\documentclass[letterpaper,twocolumn,10pt]{article}
\usepackage{usenix}
\usepackage[affil-it]{authblk} 

\microtypecontext{spacing=nonfrench}

\usepackage{colortbl}
 \usepackage{hhline}

\usepackage{enumitem}
\usepackage{cite}
\usepackage{amsmath,amssymb,amsfonts}
\usepackage{graphicx}
\usepackage{textcomp}
\usepackage[x11names]{xcolor}
\usepackage{subcaption}
\usepackage{hyperref,xurl}

\newif\ifextended
\extendedtrue

\newcommand{\ExtendedOrNormal}[2]{%
  \ifextended #1\else #2\fi
}

\ifextended
    \newcommand{\paperTitle}{InSPECtor: Improving SLEIGH Processor Specification Veracity via Proxy}
\else
    \newcommand{\paperTitle}{InSPECtor: Improving SLEIGH Processor Specification Veracity via Proxy}
\fi

\hypersetup{
    pdftitle={\paperTitle{}},
}

\usepackage{algorithm}
\usepackage{algpseudocodex}
\usepackage{newfloat}
\usepackage{multirow}
\DeclareFloatingEnvironment[fileext=lol, listname={List of Listings}, name=Listing]{listing}
\usepackage{fancyvrb}
\makeatletter
\def\verbatim@nolig@list{\do\`\do\<\do\>\do\'\do\-}
\makeatother

\CustomVerbatimCommand{\CodeInput}{VerbatimInput}{
  frame=single,
  fontsize=\footnotesize,
  commandchars=\\\{\}
}

\newcounter{recommendation}

\newcommand{\sless}[0]{\text{s<}} 
\newcommand{\slessop}[0]{\text{ s< }} 

\newcommand{\asm}[1]{\texttt{#1}}
\newcommand{\pcode}[1]{\textsc{#1}}

\newcommand{\reffig}[1]{\hyperref[#1]{Figure~\ref*{#1}}}
\newcommand{\reflisting}[1]{\hyperref[#1]{Listing~\ref*{#1}}}
\newcommand{\refsection}[1]{\hyperref[#1]{Section~\ref*{#1}}}
\newcommand{\refalgorithm}[1]{\hyperref[#1]{Algorithm~\ref*{#1}}}
\newcommand{\reftable}[1]{\hyperref[#1]{Table~\ref*{#1}}}
\newcommand{\refappendix}[1]{\hyperref[#1]{Appendix~\ref*{#1}}}
\newcommand{\refrec}[1]{\hyperref[#1]{R\ref*{#1}}}

 \newcommand{\inlinesection}[1]{\vspace{1.5px}\noindent\textbf{#1}}

\newcommand{\wip}[1]{#1}

\newcommand{\gensys}{{\textsf{GenSYS}}}
\newcommand{\jp}{{\textsf{JuxtaPlayer}}}
\newcommand{\sle}{{\textsf{SLEIGH}}}
\newcommand{\name}{{\textsf{InSPECtor}}}
\newcommand{\icicle}{{\textsf{Icicle}}}
\newcommand{\juxtaplayer}{\jp}
\newcommand{\ghidra}{\textsf{Ghidra}}

\newcommand{\intelcpu}[0]{Intel(R) Xeon(R) CPU E5-2660 v4 CPU}
\newcommand{\juxtaplayerIAcpu}[0]{AMD Ryzen 9 7950X}

\newcommand{\isacount}{5}
\newcommand{\totaldiscreps}{\hl{38,920}}

\newcommand{\xbugs}{\hl{32}} 
\newcommand{\totalbugs}{\hl{125}} 
\newcommand{\totalslefams}{35} 
\newcommand{\totalrecom}{8} 

\newcommand{\ix}[0]{{\textsf{x86-64}}}
\newcommand{\arch}[0]{\textsf{AArch64}}
\newcommand{\arm}[0]{\textsf{ARM/Thumb}}
\newcommand{\risc}[0]{\textsf{RISC-V}}
\newcommand{\msp}[0]{\textsf{MSP430}}
\newcommand{\mspx}[0]{\textsf{MSP430X}}

\newcommand*\BitAnd{\mathbin{\&}}

\usepackage{circledsteps}
\pgfkeys{/csteps/inner color=white,
        /csteps/fill color=black!50!white,
        /csteps/outer color = black!50!white, 
        /csteps/inner ysep/.initial=2pt,
        /csteps/inner xsep/.initial=3pt,
}

\usepackage{tikz}
\usepackage{xcolor}
\pgfkeys{/csteps/inner ysep=2pt}
\pgfkeys{/csteps/inner xsep=5pt}

\definecolor{snsblue}{HTML}{4C72B0}
\definecolor{snsorange}{HTML}{DD8452}
\definecolor{snsgreen}{HTML}{55A868}
\definecolor{snsred}{HTML}{C44E52}
\definecolor{snspurple}{HTML}{8172B3}
\definecolor{snsbrown}{HTML}{937860}
\definecolor{snspink}{HTML}{DA8BC3}
\definecolor{snsgrey}{HTML}{8C8C8C}
\definecolor{snsdarkgrey}{HTML}{2C2C2C}
\definecolor{snsgold}{HTML}{CCB974}
\definecolor{snscyan}{HTML}{64B5CD}

\colorlet{hlsnsblue}{snsblue!25}
\colorlet{hlsnsorange}{snsorange!25}
\colorlet{hlsnsgreen}{snsgreen!25}
\colorlet{hlsnsred}{snsred!25}
\colorlet{hlsnspurple}{snspurple!25}
\colorlet{hlsnsbrown}{snsbrown!25}
\colorlet{hlsnspink}{snspink!25}
\colorlet{hlsnsgrey}{snsgrey!25}
\colorlet{hlsnsgold}{snsgold!25}
\colorlet{hlsnscyan}{snscyan!25}

\definecolor{mplblue}{HTML}{1f77b4}
\definecolor{mplorange}{HTML}{ff7f0e}
\definecolor{mplgreen}{HTML}{2ca02c}
\definecolor{mplred}{HTML}{d62728}
\definecolor{mplpurple}{HTML}{9467bd}

\colorlet{hlmplblue}{mplblue!25}
\colorlet{hlmplorange}{mplorange!25}
\colorlet{hlmplgreen}{mplgreen!25}
\colorlet{hlmplred}{mplred!25}
\colorlet{hlmplpurple}{mplpurple!25}

\definecolor{set19c1}{HTML}{e41a1c} 
\definecolor{set19c2}{HTML}{377eb8} 
\definecolor{set19c3}{HTML}{4daf4a} 
\definecolor{set19c4}{HTML}{984ea3} 
\definecolor{set19c5}{HTML}{ff7f00} 
\definecolor{set19c6}{HTML}{ffff33} 
\definecolor{set19c7}{HTML}{a65628} 
\definecolor{set19c8}{HTML}{f781bf} 
\definecolor{set19c9}{HTML}{999999} 

\usepackage{xcolor}

\usepackage[framemethod=TikZ]{mdframed}

\usepackage[many]{tcolorbox}
\newcommand{\rqq}[1]{
 \begin{center}
     \begin{tcolorbox}[width=\columnwidth,
     colback=green!5!white,
     colframe=green!5!white,
     breakable,
     left=1pt,
     right=1pt,
     top=1pt,
     bottom=1pt,
     boxrule=0pt,         
     auto outer arc
    ]
     #1
     \end{tcolorbox}
 \end{center}
}

\newcommand{\rqqtitle}[3]{
    \phantomsection
    \refstepcounter{recommendation}
    \label{#1}
    \begin{tcolorbox}[
        enhanced,
        colbacktitle=snsdarkgrey!20!white, 
        left=1pt,
        right=1pt,
        top=1pt,
        bottom=1pt,
        boxsep=2pt,
        arc=2pt,
        fonttitle=\bfseries,
        frame hidden,
        title={\textcolor{black!50!black}{\#\arabic{recommendation}:~#2}},
        boxrule=1pt,
        titlerule=1pt,
        titlerule style=black!50!black,
        ]
        #3
     \end{tcolorbox}
}

\newcommand{\hl}[1]{\textcolor{black}{#1}} 

\ifextended
\else
\usepackage[available]{usenixbadges}
\fi

\begin{document}

\date{}

\title{\Large {\bf \paperTitle{}}}

\author[1]{Michael Chesser}
\author[2]{Paul Quirk}
\author[1]{Douglas Cooke}
\author[1]{Guy Farrelly}
\author[3]{Surya Nepal}
\author[1]{Damith C. Ranasinghe}

\affil[1]{Adelaide University}
\affil[2]{Defence Science and Technology Group}
\affil[3]{CSIRO}

\maketitle

\begin{abstract}
Processor specifications underpin critical security and program‑analysis tools such as disassemblers, decompilers, and emulators, yet, their correctness is rarely examined. Errors in specifications distort program behavior, obscure vulnerabilities, and enable analysis‑evasion techniques. Validating processor specifications is a non-trivial task. 
Our study is a significant undertaking to enable the systematic validation of open-source \sle{} language specifications, predominantly used by \ghidra{}. We design and implement a testing framework based on an automated oracle validation strategy by \textit{proxy}. Our approach leverages the structure encoded in a specification itself to enumerate decodable instruction forms and generate targeted initial states. Then differentially test the successful decoding and emulation of those instructions by comparing emulators exercising the processor specification against hardware references. 

Applying \name{} across diverse, open-source specifications---\ix{}, \arch{}, \arm{}, \risc{}, \msp{}---embedding differences in specification styles, author preferences,  and instruction set architecture designs, we uncovered over \totaldiscreps{} discrepancies that led to 125 unique bugs with proposed fixes, identifying decoding and semantic defects as well as cross‑vendor inconsistencies. We distill our findings into \totalrecom{} concrete recommendations to drive future improvements. Our work underscores the importance of specification correctness and provides a practical tool to substantially improve the fidelity of \sle{} processor specifications, strengthening the reliability of downstream security and analysis tools.
\end{abstract}

\section{Introduction}
\sle{} is a domain-specific language (DSL) for describing processor specifications that decode and translate instructions into the architecture-agnostic intermediate representation (IR), Pcode~\cite{sleighreference2023}. There are over \wip{\totalslefams{}} processor families expressed in \sle{} specifications, covering a diverse range of processors from \textsf{PIC} microcontrollers to modern \ix{} and \arch{} CPUs. Significantly, \sle{} specifications underpin Ghidra~\cite{ghidra2019}, the dominant open source multi-architecture toolchain for reverse engineering and binary analysis, and multi-architecture emulators such as Alligator from Airbus~\cite{ghidralligator2023}, Icicle~\cite{icicle2023}, MetaEmu~\cite{metaemu2022} and Styx~\cite{styx2025}. In Ghidra’s disassembler and decompiler, \sle{} specifications directly determine how binary code is decoded, lifted, and interpreted by analysts and automated scripts. As a result, \sle{} specifications sit on the critical path of security workflows such as vulnerability discovery, malware analysis, fuzzing, firmware auditing, and program understanding.

This widespread reliance amplifies the consequences of specification errors: flaws in \sle{} can propagate across multiple tools, users, and workflows. Yet, despite this central role, the reliability of \sle{} specifications remains an underexplored problem.

In practice, \sle{} specifications are large, painstakingly constructed artifacts that encode both instruction formats and their semantics. Developing faithful processor specifications is a non-trivial task. For instance, the ARM manual is 12,000 pages where complexities such as the operand-dependent behavior of instruction prevents a simple approach based on examining mnemonics. Additionally, the 2025 edition of the Intel Software Developer's Manual~\cite{IntelMaual2025} exceeds 5,000 pages. The specifications must capture subtle architectural behaviors across diverse and often poorly documented instruction set architectures. Further, to address the demand for rich, fast, and safe computation, modern architectures are constantly evolving, with new instructions and features added each year\footnote{For instance, microprocessor advancements, such as Single Instruction, Multiple Data (SIMD), Arm Neon, Advanced Matrix Extensions (AMX), Advanced Vector Extensions (AVX) or cryptography extensions are often incorporated to cater to artificial intelligence, high-performance computing and security-focused workloads.}, translating to an increase in instructions and complex instruction-bit encodings in ISAs. The painstaking construction and update of specifications to integrate architectural changes while maintaining backward compatibility makes the specifications intrinsically susceptible to bugs. Even small mistakes---such as missing side effects or incorrect semantic ordering---can propagate silently through static and dynamic analysis tools. These errors may lead to misleading decompilation output and unsound program analyses, ultimately undermining the conclusions drawn by security practitioners; often exploited by malware to detect virtual environments~\cite{avleak2016} and hide from analysis~\cite{redpills2009,cardinalpill2014,handlingantivm2017}.

Importantly, the critical dependency on \sle{}  specifications has not been met by a commensurate, open source assurance of its correctness. Therefore, we are motivated to improve the correctness of \sle{} processor specifications.

\rqq{By enhancing the veracity of \sle{} processor specifications, we expect to drive improvements across all their applications.}

\inlinesection{The Study.~}In this \textit{first} study, we focus our efforts on improving the veracity of the \sle{} processor specifications with the hope of generalizing our learning outcomes to broader specification assurance research. Importantly, these specifications need to be updated and reassessed for correctness as ISAs evolve and new instructions are added. Our efforts will facilitate testing implementations thoroughly, and automating this process in the future. 

Previously, studies have examined improving the correctness of CPU emulators~\cite{emufuzzer2009, pokeemu2012, kemufuzz2010, fastpokeemu2018}, hypervisors~\cite{virtcpuvalidation2015}, processor hardware~\cite{difuzzrtl2021, cascade2024, silifuzz2021}, disassembly veracity (textual representation accuracy)~\cite{nversiondisasm2010, examiner2022}, while a few have tested the semantic correctness of binary lifters~\cite{testIR-ASE-kim2017,McSemaTesting-PLDI-Dasgupta2020} but specification-level correctness has received less attention. Although the work in~\cite{naus2022} considered usability of Pcode for formal analysis and the authors in~\cite{ASL4ISAs-Alistair2019} made significant strides to validate the veracity of Sail specification language described semantic models for significant parts of \textsf{ARMv8-A}, \risc{}, and \textsf{CHERI-MIPS}, the correctness of the extensive multi-architecture support within the \sle{} specifications has not been examined.

In seeking to design an approach to uncover specification bugs, we observe that a processor specification itself encodes enough structure to drive systematic validation. So, we introduce \name{}, a framework that automatically traverses \sle{} specifications to enumerate decodable instructions, derive targeted execution states from semantic definitions, and perform differential testing against hardware reference systems.  This enables high‑coverage, specification‑driven testing without redundant instruction exploration.

\vspace{1mm}
\noindent\textbf{Our Contributions.~}In summary:
\begin{itemize}[itemsep=2pt,parsep=1pt,topsep=2pt,labelindent=5pt,leftmargin=12pt]
  \item We propose symbolically traversing \sle{} specification decoding rules, a new, systematic methodology that aims to exhaustively generate instruction encodings, extracting edge cases for testing and deriving initial input states that trigger these edge cases.
  \item We design, implement, and evaluate the methodology in the prototype \name{} system for testing specifications, including, \gensys{}---a test case generation tool that integrates these techniques and \juxtaplayer{}---a robust differential testing environment capable of executing test cases on both emulators and hardware-assisted virtual machines.
  \item We analyze the discovered bugs, identify their root causes, propose fixes, and distill \totalrecom{} recommendations for improving the \sle{} DSL and preventing future issues.
\end{itemize}

\rqq{
Importantly, we use \name{} to test \ix{}, \arch{}, \arm{}, \risc{}, and \msp{} \sle{} specifications. Our work uncovered over \wip{\totaldiscreps{}} discrepancies. Manually triaging, categorizing, and prioritizing the discrepancies, we identified the root causes of the discrepancies leading to the discovery of \totalbugs{} unique bugs with proposed fixes submitted to \ghidra{} maintainers. We open-source \name{} on GitHub: \href{https://github.com/Sleigh-InSPECtor/}{https://github.com/Sleigh-InSPECtor}.
}

\section{Instruction Encodings and \sle{} Specifications Primer}\label{sec:emu-testing:background}

\begin{figure*}[t]
  \centering
  \includegraphics[width=0.80\textwidth]{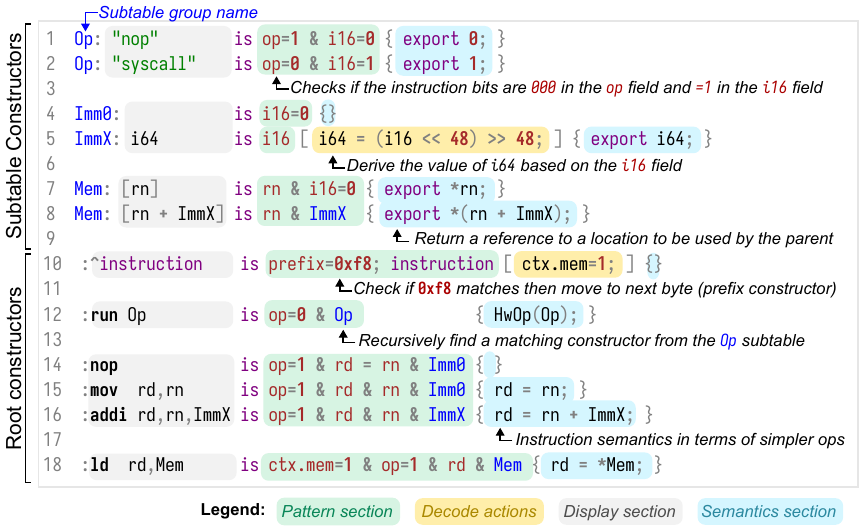}
  \caption{An example of a \sle{} specification for a simple ISA with a variable length instruction set, such as \ix{}.}\label{fig:emu-testing:sleigh-overview}
\end{figure*}

Before describing our specification-driven testing approach, we briefly review ISA-level instruction encodings and the structure of \sle{} specifications. Throughout this discussion, we use examples from a simplified ISA with a variable length instruction set, similar to \ix{}, defined by the \sle{} specification in \reffig{fig:emu-testing:sleigh-overview}\footnote{For simplicity, we omit the parts of the specification that define the register and memory configuration of the ISA, as well as the mapping from field names to bit ranges in the instruction encoding.}.

Instructions are represented as sequences of bytes, referred to as \textit{instruction encodings}. Although each concrete instruction has a unique binary representation, instruction encodings within an ISA typically share a common structure. Bits are grouped together to form fields that have specific meanings, such as the registers, immediate values, and addressing modes. For example, in the simplified ISA, \texttt{addi r2,r3,0x1234} is encoded as follows:

\begin{figure}[H]
\centering
    \includegraphics[width=0.88\linewidth]{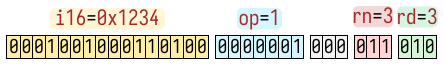}
\end{figure}

\sle{} specifications exploit this structure by expressing decoding as a hierarchy of declarative rules. These rules, called \textit{constructors}, are organized into groups called tables. The \sle{} specification in \reffig{fig:emu-testing:sleigh-overview} is annotated with the key components of each constructor:

\begin{itemize}[itemsep=2pt,parsep=1pt,topsep=3pt,labelindent=5pt,leftmargin=12pt]
    \item \textbf{Pattern section}:~defines the mapping from an instruction encoding to the other components of the specification via Boolean expressions that reference the instruction encoding and the current decoder \textit{context}---a \sle{}-specific mechanism  to encode or transfer information between constructors which we will discuss later.
    \item \textbf{Decode actions}:~optional expressions, after the pattern section between `[ ]', that specify how to update the decoder context when the constructor is evaluated.
    \item \textbf{Display section}:~defines the textual representation of the instruction used for disassembly. 
    \item \textbf{Semantic section}:~defines the meaning of the instruction in terms of the simpler ISA-independent operations (Pcode).
\end{itemize}

\noindent
Subtable constructors that share a group name (e.g., \texttt{Op}, \texttt{Mem}) form a table and are responsible for decoding a particular aspect of the instruction, such as operand decoding, and are used at different points during the decoding process.
Constructors that do not specify a table name (i.e. those that start with `\texttt{:}'), are implicitly associated with the \textit{root} table, and are the starting point for the decoding process. All other tables are used for decoding only when they appear in the pattern section of one of the root-level constructors (such as \texttt{Op} on Line 12).

Employing a \sle{} specification for decoding begins with two inputs: a sequence of bytes (the instruction encoding) and a set of context bits. Context bits are a \sle{}-specific mechanism used for two purposes: i)~to encode information relevant to decoding that is not directly present in the instruction encoding, such as the current processor mode (e.g., \textsf{ARM} vs. \textsf{Thumb}); and ii)~to transfer information between constructors. For example, context bits are often used to alter operand decoding in the presence of prefix bytes.

Given this initial state, the decoder attempts to find a constructor whose \textit{pattern section} matches the current state. Constructors are considered from most to least specific, where specificity is determined by the number of constrained bits in the pattern section\footnote{Tie-breaking depends on the declaration order, and non-overlapping constructors can be matched out-of-order.}. If no constructor matches, the encoding is treated as undecodable or invalid. When a match is found, the decoder applies the corresponding decode actions to update its state. It then recursively evaluates any subtables referenced in the \textit{pattern section} using the updated state. Constructors in each subtable are matched and evaluated in the same way until the instruction is fully decoded. Once decoding is complete, the instruction semantics can be determined by evaluating the \textit{semantic section} of the constructors matched during decoding, potentially substituting subtable references with the values exported by the corresponding constructors.

\begin{figure*}[t]
  \centering
  \includegraphics[width=0.95\textwidth]{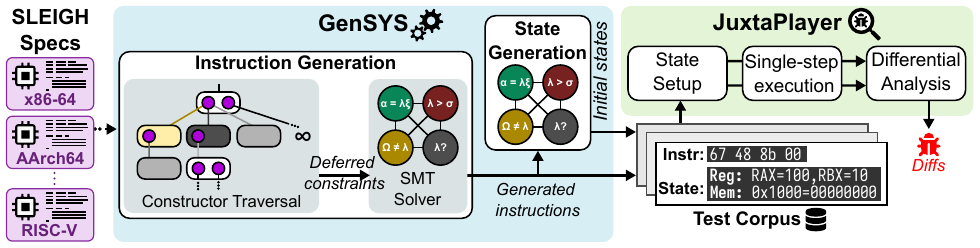}
  \caption{A high-level overview of \name{}; the specification-driven test case generation (\gensys{} detailed in \refsection{sec:ins-gen} \& \ref{sec:emu-testing:state-generation}) and differential testing (\juxtaplayer{} detailed in \refsection{sec:emu-testing:execution-environment}) process to identify bugs or defects in the specification encodings.
}
  \label{fig:emu-testing:overview}
\end{figure*}

\textbf{\textit{Example.}~}Consider the \texttt{addi r2,r3,0x1234} instruction from before. After extracting the fields from the instruction encoding, the decoder sees: \texttt{rd=2}, \texttt{rn=3}, \texttt{op=1}, \texttt{i16=0x1234}, and attempts to find a constructor in the root table matching these values. 
The \texttt{nop}, \texttt{mov} and \texttt{addi} constructors all match the current opcode (\texttt{op=1}). However, \texttt{nop} is rejected because \texttt{rd$\ne$rn} and \texttt{mov} is rejected because \texttt{i16$\ne$0}\footnote{This requires evaluating the \texttt{Imm0} constructor and \textit{backtracking} which is described in more depth in \refsection{constraints}.} causing the \texttt{addi} constructor to match.

The decoder then evaluates any decode actions (none for the \texttt{addi} constructor) and recurses into any referenced subtables. The \texttt{addi} constructor has a single referenced subtable (\texttt{ImmX}) which has one constructor (Line~5). This constructor has no additional constraints (the \texttt{i16} field is unconstrained), so it matches immediately, and the decode action computes \texttt{i64=0x1234}. At this point, all operands have been decoded\footnote{Mapping register fields (\texttt{rn} and \texttt{rd}) to specific registers is handled by attachments, which is described in \refsection{constraints}.}, and the target disassembly is obtained by substituting these decoded values into the display section of the matched constructor, yielding \texttt{addi r2,r3,0x1234}. Similarly, the instruction semantics are obtained by substituting the decoded operands into the semantic section, resulting in the instruction semantics being: \texttt{r2=r3+0x1234}. For a complete \sle{} DSL description, readers are directed to the resource in~\cite{sleighreference2023}. 

For instruction generation, we are interested in traversing all possible decode paths which requires considering the pattern and decode sections of the constructors we discussed in this section.

\section{Methodology}

\reffig{fig:emu-testing:overview} presents an overview of our approach, \gensys{} systematically generates decodable instructions as test cases, effectively aiming to cover all instructions recognized by a \sle{} specification; \juxtaplayer{} applies a differential testing strategy, comparing the behavior of executing an instruction in a \sle{}-based emulator with a hardware reference. The key idea is to employ the \sle{}-based emulator as a proxy for assessing the correctness of a target specification.

Ideally we would systematically generate and test every unique instruction recognized by a \sle{} specification. However, testing every possible combination of operand and immediate values is both computationally wasteful and unnecessary, since most operand variations do not affect an instruction's core behavior. Instead, we generate all instructions with unique semantics by traversing a specification's constructors as we detail in \refsection{sec:ins-gen}, and combine them with a representative subset of operand values: simple defaults, random selections, and values that force operand aliasing within the instruction semantics. In the following, we elaborate on the instruction generation and state generation methods in \gensys{} followed by the testing method encapsulated in \juxtaplayer{}.

\subsection{Instruction Generation}\label{sec:ins-gen}


For \sle{} specifications, generating all instructions with unique semantics is equivalent to generating one instruction for each possible decode path in the specification. We can do this by determining which constraints on the instruction encoding must hold for a given constructor to be selected during decoding, and then solving those constraints to produce a concrete instruction encoding.

Conceptually, this process can be viewed as a constraint satisfaction problem (CSP). In practice, however, efficiently collecting and solving all relevant constraints requires a complex traversal of the decode tree. In particular, the traversal must handle several kinds of constraints, account for implicit constraints introduced by overlapping constructors, manage decode actions that modify decoder state during traversal, and avoid unbounded recursion. 

The core of the traversal process is outlined in \refalgorithm{algorithm:emu-testing:get-instructions}. 
Starting from the root-level table, {\sc GetInstructions} invokes the {\sc ConstructorTraversal} algorithm, which selects a constructor $c$ for the current subtable $t$ (forking the current state $S$ if there were multiple valid choices), translates all constraints required to match that constructor to symbolic expressions ({\sc Translate}), updating $S$, before recursively repeating the process for any subtables. The state returned by {\sc ConstructorTraversal} contains the complete set of constraints associated with the decode path, which are solved using an SMT solver to produce concrete instruction encoding $X$. {\sc GetInstructions} then restores the state to a previous fork and selects a different constructor, repeating this process until every possible path has been explored.

The algorithm's state $S$ consists of three components: i)~an instruction bit vector $\mathbf{I} = b_1b_2\cdots b_n$; ii)~a context bit vector $\mathbf{C} = b_1b_2\cdots b_m$; and iii)~a collection of deferred constraints. For the bit vectors, each bit is either a fixed value ($0$ or $1$) or symbolic ($?$). The size of the context bit vector ($m$) is a constant defined by the specification. For variable-length ISAs, the size of the instruction bit vector ($n$) is not known until traversal is complete, instead we define an upper bound\footnote{Considering the largest possible instruction across all tested ISAs is 120 bits (15-byte x86-64 instructions), $\text{max}(n) = 120$.} and ignore any unreferenced bits after all constructors have been selected.


\noindent\scalebox{0.77}{%
\begin{minipage}{1.0\linewidth}
\begin{algorithm}[H]
\begin{algorithmic}[1]
\Function{GetInstructions}{}

  \State $O \gets \{\}$ \Comment{The list of generated instructions (Output).}
  \State $F \gets \{\}$ \Comment{The set of fork points for unselected paths.}

  \State $S \gets $ \textit{initial state}, $t \gets $ \textit{root subtable}
  \While{$S \ne \varnothing$}
    \State $S, f \gets$ \Call{ConstructorTraversal}{$S$, $t$}
    \State $F \gets F \cup f$ \Comment{Add new fork points found during subtable traversal.}

    \State $X \gets$ \Call{Solve}{$S$} \Comment{Get instruction bytes satisfying all constraints.}
    \If{$X$ \textbf{is} \textit{SAT}}
      \State $O$.push($X$)
    \EndIf

    \State $(S, t) \gets F.\text{pop}()$
  \EndWhile

  \State \Return $O$
\EndFunction
\\
\State {$t$: subtable to add constraints from}
\State {$S$: the current decoder state}
\Function{ConstructorTraversal}{$S$, $t$}
    \State $\text{F} \gets \{\}$

    \If{$S$ \textbf{at} \textit{traversal limit}}
      \State \Return (\textit{UNSAT}, $F$)
    \EndIf

    \State $c \gets$ \Call{CurrentConstructor}{$S$, $t$}

    \If{$S$ \textbf{has} \textit{unselected constructors}}
      \State $F$.push(\Call{NextConstructor}{$S$, $t$})
    \EndIf

    \State $S.\text{simple } \gets S.\text{simple } \land $ \Call{Translate}{$c.\text{simple}$}

    \If{$S$.simple \textbf{is} \textit{UNSAT}}
      \State \Return (\textit{UNSAT}, $F$) \Comment{Terminate traversal early if simple constraints are UNSAT.}
    \EndIf

    \State $S.\text{deferred } \gets S.\text{deferred } \land \Call{Translate}{c.\text{cplx}} \land \Call{Translate}{c.\text{attach}}$

    \For{$o_i$ \textbf{in} \Call{FindOverlaps}{$S$, $c$}}
      \State $S.\text{deferred} \gets S.\text{deferred } \land \neg\Call{Translate}{o_i}$
    \EndFor

    \State $S \gets $ \Call{ApplyDecodeActions}{$S$, $c$}

    \For{$t_i$ \textbf{in} \Call{Subtables}{$c$}} \Comment{Recursively traverse subtables and merge constraints.}
      \State $S, f \gets$ \Call{ConstructorTraversal}{$S$, $t_i$}
      \State $F \gets F \cup f$
      \If{$S$ \textbf{is} \textit{UNSAT}}
        \State \Return (\textit{UNSAT}, $F$)
      \EndIf
    \EndFor

    \State \Return ($S$, $F$)
  \EndFunction
\end{algorithmic}
\caption{
Instruction Generation}\label{algorithm:emu-testing:get-instructions}
\end{algorithm}
\end{minipage}
}
\vspace{2mm}


Every bit in the instruction bit vector is initialized as symbolic ($\mathbf{I}_i = \text{ ? for }i \in { 1, 2, \cdots n }$), while the context bit vector is initialized to a fixed value determined by the current processor configuration (for our simple ISA specification in \reffig{fig:emu-testing:sleigh-overview}, it is initialized to 0).

Deferred constraints are additional constraints on the instruction bit vector that cannot be resolved to a fixed value (e.g., inequalities). These constraints are solved simultaneously after the traversal process has completed and all constraints are known. By maintaining known instruction bits separately from deferred constraints, the algorithm can efficiently test satisfiability during traversal and prune large portions of the search space. For example, if a constructor in a subtable requires a particular bit to be 1, but that bit is already known to be 0 from an earlier constraint, then the corresponding subtree can be discarded immediately, significantly improving efficiency. Next, we detail the complex component of the method  employed in the instruction generation algorithm design.

\subsubsection{Translating Constructor Constraints}\label{constraints}
The following section describes the method for translating the pattern section of each constructor into a combination of fixed bit patterns and deferred constraints encapsulating the {\sc Translate} function in \refalgorithm{algorithm:emu-testing:get-instructions}.

\inlinesection{Simple constraints.} Most constraints involve equality checks between a constant and a range of instruction or context bits. These constraints can be translated directly into fixed values for the corresponding bits in the instruction or context bit vector. If a new constraint is unsatisfiable given the previously fixed values, traversal stops and the algorithm continues with a different set of constructor choices. Otherwise, the newly constrained bits are updated and traversal continues.

\textbf{\textit{Example.}~}Using the specification \reffig{fig:emu-testing:sleigh-overview}, consider the \texttt{:run} constructor (Line 12) and the \texttt{Op} subtable (Lines 1--2):

\begin{figure}[H]
    \includegraphics[width=0.90\linewidth]{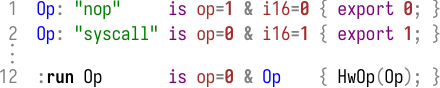}
\end{figure}
When the \texttt{:run} constructor is selected, the algorithm updates the instruction bit vector so that the \texttt{op} field bits are fixed to 0. It then attempts to select a constructor from the \texttt{Op} subtable by checking each constructor's constraints against the current state. The \texttt{Op:"nop"} constructor requires the \texttt{op} field is equal to 1, which conflicts with the current value of 0, so it is immediately discarded.
The \texttt{Op:"syscall"} constructor is compatible, so it is selected and traversal continues after adding a \asm{i16=1} constraint.

All simple constraints can be reduced to a bitwise mask-and-compare operation. For example, when generating an instruction corresponding to \texttt{run syscall}, the final symbolic expression passed to the SMT solver is $(\mathbf{I} \BitAnd \mathtt{0xfffffe00} = \mathtt{0x00010000})$, where $\mathbf{I}$ denotes the instruction bit vector.

\inlinesection{Complex constraints.} Beyond simple equality constraints, \sle{} also allows constructors to use more complex constraints to match instructions. This includes both inequalities and constraints involving non-constant expressions. Unlike simple constraints, there is often no single value that can be assigned to the constrained bits to satisfy the constraint outright. One possible strategy would be to choose \emph{any} value satisfying the constraint and continue traversal; however, such a choice may conflict with constraints introduced by later subtables, preventing valid instructions from being generated.
Another option would be to exhaustively explore all possible values, but this would require traversing an excessive number of paths and would generate many redundant instructions.
Instead, we elect to check that such constraints are \emph{potentially} satisfiable (i.e., they do not already conflict with the existing simple constraints) and then add them to a list of deferred constraints, which are solved once traversal is complete.

Once constructors for all subtables have been selected, the full set of constraints for the current instruction is known. To obtain a final instruction encoding, a symbolic expression is constructed from the instruction bit vector together with the deferred constraints. This expression is then passed to an SMT solver (Z3), which attempts to find a concrete assignment to all bits satisfying the combined constraints. If the solver succeeds, it returns a concrete instruction encoding; otherwise, it reports that the constraints are unsatisfiable.

\textbf{\textit{Example.}~}As an example, consider the \texttt{:nop} constructor from Line~14 of our example specification in \reffig{fig:emu-testing:sleigh-overview}:
\begin{figure}[H]
    \includegraphics[width=0.7\linewidth]{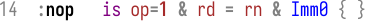}
\end{figure}
For an instruction to match this constructor, the \texttt{rd} and \texttt{rn} fields must be equal. This constraint cannot be translated directly into fixed values for the instruction bit vector, since there are many possible values for \texttt{rd} and \texttt{rn} that satisfy it. Rather than choosing an assignment prematurely, we add the constraint \texttt{rd=rn} to the list of deferred constraints. At the end of traversal, the final expression passed to the SMT solver is:%
\[
    (\textbf{I} \BitAnd \mathtt{0xfffffe00} = \mathtt{0x00000200}) \land (\textbf{I}.rd = \textbf{I}.rn)
\] %

To optimize the process further, we check whether deferred constraints remain potentially satisfiable given the currently known bits. Suppose there is a deferred constraint \texttt{i16 > 0}. The algorithm then checks whether at least one bit in the range corresponding to \texttt{i16} is either unknown or equal to 1. If a different constructor requires that \texttt{i16=0}, we can immediately reject it due to the \texttt{i16 > 0} constraint, which avoids wasting time traversing impossible constructor combinations.

\inlinesection{Constraints imposed in attachments.} Fields referenced by a constructor may be linked to attachments that control the interpretation of the underlying bits. While most attachments allow all field values, some \sle{} specifications use attachments to mark certain operand values as invalid. To handle this during instruction generation, invalid values are excluded by introducing appropriate deferred constraints.

\textbf{\textit{Example.}~}A possible set of attachments for our example specification in \reffig{fig:emu-testing:sleigh-overview} is as follows:

\begin{figure}[H]
    \includegraphics[width=0.90\linewidth]{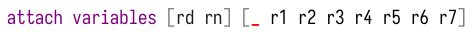}
\end{figure}%
Whenever \texttt{rd} or \texttt{rn} appears unconstrained in the specification (e.g., in the \asm{:mov} and \asm{:addi} constructors), its value is interpreted according to the attachment mapping. For example, \texttt{rd=1} maps to \asm{r1}; \texttt{rd=2} maps to \asm{r2}, and so forth. In this example, an attachment index of 0 has no associated register, indicated by the `\texttt{\_}' operator. To account for this, we treat such cases as implicit complex constraints: \texttt{rd$\ne$0} and \texttt{rn$\ne$0}. For instruction generation, these constraints are handled in the same way as other complex constraints discussed previously.

\inlinesection{Implied constraints from overlapping constructors.} The behavior of an instruction is primarily determined by its `opcode' field,\footnote{To keep instruction encodings compact, the full opcode may be split over multiple fields depending on the ISA.} which specifies the operation performed on the instruction's operands. In general, two instructions with the same opcode have the same semantics regardless of their operand values. However, some instructions can be represented as a specific sub-encoding of a more general instruction. For example, copying a value between registers (e.g., \asm{mov r0,r1}) is semantically equivalent to adding zero to a source and storing the result in a destination register (e.g., \asm{addi r0,r1,0}). These instructions are often referred to as \textit{pseudo-instructions}, and are present in many ISAs.

Although \sle{} specifications could ignore pseudo-instructions and always decode them as their generic counterparts, modeling them explicitly benefits both human and machine analysis. A more specific textual representation improves disassembly for human analysts, and defining specialized semantics can simplify both static and dynamic analysis.

In addition to pseudo-instructions, some ISAs include instructions with more complex operand encodings that require special handling during decoding. For example, in both \ix{} and \msp{}, instruction encodings include `addressing mode' bits that alter the interpretation of operand fields. 

To model such encodings, \sle{} provides a mechanism known as constructor specialization, which allows multiple constructors within the same subtable to match the same instruction encoding. Recall (\refsection{sec:emu-testing:background}), the \sle{} decoder prioritizes matching constructors from most to least constrained. Consequently, for an instruction to match a less constrained constructor, it must also avoid matching any overlapping constructors that would be considered first. To model this behavior, we treat the target constructor as if it carried an additional implied constraint requiring all preceding overlapping constructors to not match. 

\textbf{\textit{Example.}~}Consider Lines~14--16 of \reffig{fig:emu-testing:sleigh-overview}:
\begin{figure}[H]
    \includegraphics[width=\linewidth]{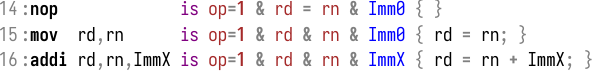}
\end{figure} %
To generate a \asm{mov} instruction, we must avoid matching the \asm{:nop} constructor. Since the only difference between the two constructors is the constraint \texttt{rd$=$rn}, we add its negation, \texttt{rd$\ne$rn}, to the list of deferred constraints for the \asm{:mov} constructor.

In rare cases, there are no unique constraints that can be added to distinguish between two overlapping constructors. For example, consider the \texttt{:addi} constructor in the example specification. This constructor overlaps with both \texttt{:nop} and \texttt{:mov}. Similarly, we can avoid matching \texttt{:nop} by adding the constraint \texttt{rd$\ne$rn}. However, at the constructor level, the constraints for \texttt{:mov} and \texttt{:addi} are identical.

The only difference between the two constructors is the final subtable (i.e., \texttt{Imm0} and \texttt{ImmX}). Since subtables are evaluated only after a constructor has been selected, it may seem impossible for \texttt{:addi} to match. However, \sle{} supports \textit{backtracking}; if no constructor matches in a later subtable, the decoder may return to an earlier point in the process and try a different constructor.

Therefore, to ensure that \texttt{:addi} is selected, we must force the \sle{} decoder to backtrack from the \texttt{:mov} constructor. To achieve this, the generation algorithm inspects all constructors in the \texttt{Imm0} subtable and adds additional constraints that ensure that no constructor matches. In this example, the \texttt{Imm0} constructor requires \texttt{i16$=$0}, so we add the constraint \texttt{i16$\ne$0}. If the subtable contained multiple constructors, we would instead need a combined constraint ensuring that all of them fail to match.
The final set of constraints the algorithm solves for the \texttt{:addi} constructor is:

%
%
\begin{figure}[H]
    \includegraphics[width=0.9\linewidth]{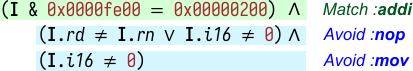}
\end{figure}%

Notice that when we add constraints to resolve the overlap with the \texttt{:nop} constructor,  the overlap can be resolved by negating the simple constraints \textit{or} by backtracking, maximizing the flexibility the solver has to generate an encoding, allowing \texttt{addi r0,r0,1} to be a valid solution.


\inlinesection{Self-referencing constructors and decode actions.}
\sle{} allows constructors to be \textit{self-referencing}, meaning that a constructor may reference the same subtable to which it belongs. This functionality is used for two main purposes: i)~\textit{phased decoding} and ii)~optional \textit{prefix parsing}.
In both cases, the self-referencing constructor must be able to influence subsequent steps of the decoding process. To support this, constructors may specify \textit{decode actions} in square brackets after the pattern section. Each decode action evaluates an arithmetic expression, possibly over context and/or instruction bits, and writes the result to a context field.

\emph{Phased decoding.} For phased decoding, context fields are used to split the decoding process into multiple phases, where previous phases manipulate other context bits which are used as constraints in later phases.

During instruction generation, this mechanism is straightforward to handle; at any point in the traversal, the phase has a fixed value, so only constructors for the current phase need to be considered. Phased decoding is not used in our example specification, instead we provide the following example (inspired by the \arch{} specification) to explain.

\textbf{\textit{Example.}~}Consider the case where the immediate field is split across the instruction encoding:



\begin{figure}[H]
    \includegraphics[width=\linewidth]{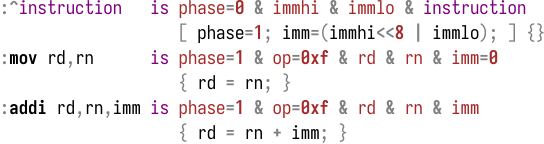}
\end{figure}

Assuming the \texttt{phase} context field is initialized to 0, only the first constructor matches the current context. The instruction generation algorithm therefore selects it and updates the context according to the decode actions. When the algorithm subsequently resolves the \texttt{instruction}\footnote{Recall that the \texttt{instruction} identifier is a special identifier that matches the root subtable---constructors prefixed with just `:'.} subtable, it uses the updated context, where \texttt{phase$=$1}. Since the phase value is now 1, only the second and third constructors match, allowing them to act as if they were separate subtables. The \texttt{:mov} constructor demonstrates the usefulness of phased decoding, where it is able to define a constraint on the entire immediate value instead of separate constraints on each subfield. 

\emph{Prefix parsing.~}Constructors used for prefix parsing are more involved to translate. Since prefix constructors also consume part of the instruction bit vector, the algorithm must track the offset within the instruction encoding, and adjust any subsequent constraints. 
Additionally, since prefix bytes can be optional and repeated (e.g., \ix{}), unlike phased parsing, prefix constructors cannot be easily separated from regular opcode decoding, which can lead to implied constraints from overlap resolution. Further, to avoid generating arbitrarily long encodings from repeated prefixes, we must incorporate a traversal limit during the traversal process.\footnote{Given that physical hardware will always have an upper limit on instruction length, this should never prevent us from generating a valid instruction.}

\textbf{\textit{Example.}~}Consider the \texttt{:ld} constructor:
\begin{figure}[H]
    \includegraphics[width=\linewidth]{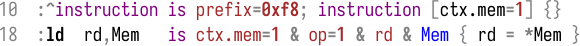}
\end{figure}
For the \texttt{:ld} constructor to match, instead of one of the arithmetic instructions with the same opcode, the context field \texttt{ctx.mem} must be set to 1. This occurs only if the \textit{prefix} constructor (Line~10) is matched first; that constructor adds the constraint that the instruction must begin with the prefix byte (\texttt{0xf8}) and updates the context field \texttt{ctx.mem} to 1. Consequently, when the instruction generation algorithm tries to generate an encoding for the \texttt{:ld} constructor, it emits the following constraint:\footnote{Constraints simplified; in practice, additional constraints are required for overlap resolution and attachments.} $\mathbf{I} \BitAnd \mathtt{0x0000fe00ff} \ne \mathtt{0x00000400f8}$.

\emph{Dynamic context.~}If the result of a decode action depends on unconstrained parts of the instruction encoding, the corresponding parts of the context bit vector become partially symbolic. Because the only source of symbolic values is the instruction bit vector, each value can be expressed as a function of instruction bits and constants. Any later constraint on those context bits can therefore be translated into a constraint on the instruction bit vector. These translated constraints are added to the deferred-constraint set and solved after traversal.

\textbf{\textit{Example.}~}Consider the following extension to our example specification:


\begin{figure}[H]
    \includegraphics[width=0.9\linewidth]{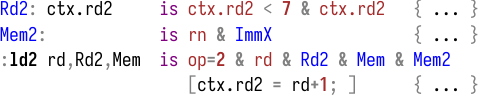}
\end{figure}

The instruction generation algorithm must track that \texttt{ctx.rd1} is equivalent to \texttt{rd+1}, and ensure that a subsequent constraint on \texttt{ctx.rd1} is mapped to a corresponding constraint on \texttt{rd}. For example, the \texttt{ctx.rd1 < 7} constraint in the \texttt{Rd2} subtable becomes \texttt{rd+1 < 7}.

\subsubsection{Avoiding Excess Constructor Combinations}

\begin{figure}[h]
    \centering
    \includegraphics[width=\linewidth]{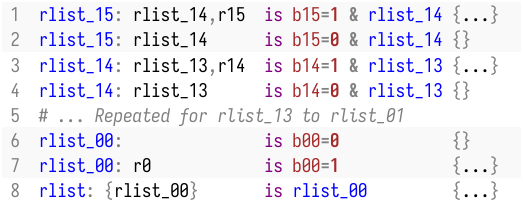}
    \caption{Register list operand decoding in the ARM \sle{} specification (e.g., for \texttt{LDMDA r0, \{r1,r2 ...\}}): Each register in the list can be either present or absent and there are 16 registers (\asm{r0} to \asm{r15}) leading to $2^{16}$ combinations.}\label{arm-reglists}
\end{figure}

\sle{} specifications use attachments to map the numerical value of a field to a specific register. However, more complex relationships between fields which cannot be easily expressed with attachments are sometimes required. In such cases, specifications resort to mapping field values through separate constructors, which exponentially increases the number of constructor combinations. For example, ARM allows any combination of the registers \asm{r0} to \asm{r15} to appear in the register lists used by the load/store-multiple instructions \asm{LDMIA} and \asm{STMIA}. The corresponding decoding specification (\reffig{arm-reglists}) requires traversing approximately $2^{16}$ paths to cover all operand combinations. To avoid generating an instruction for every combination, the algorithm tracks which constructors have already been covered and prunes them during subtable traversal when this type of explosion is detected (\textit{traversal limit} on Line~18).

\subsubsection{Solving Instruction Encodings}

After returning from {\sc ConstructorTraversal}, all constructors have been selected. However, when we attempt to solve the constraints to find a concrete instruction encoding (Line~8), there can still be several bits in the encoding that remain unconstrained, or only partially constrained (e.g., \asm{imm > 0}). These `free' bits correspond either to generic operands or to unused bits that do not affect the instruction. As a result, each instruction typically has a range of valid encodings. Exhaustively generating and testing every combination of these free bits would reintroduce the same exponential explosion we are trying to avoid. Instead, the algorithm generates a smaller representative set of solutions that covers the most \textit{interesting} cases.

To do this, {\sc Solve} attempts to identify possible edge cases in the instruction encoding. It uses Z3's support for minimizing and maximizing values~\cite{nuz2015} to generate test cases at the minimum and maximum encodings permitted by the constraints, which helps expose potential off-by-one errors in partial constraints. For example, for the \texttt{:nop} constructor (Line 14 of our example specification), given the complex constraint \texttt{rd=rn}, Z3 generates encodings corresponding to \texttt{rd=rn=0} and \texttt{rd=rn=7}.

Next, analysis of the instruction's semantics identifies any registers that could alias with operands derived from the unconstrained bits. Additional constraints are then added to the instruction bit vector to force operands to match the aliasing register, generating targeted cases for each potential alias.

\begin{table*}[t]
    \centering
    \resizebox{0.95\linewidth}{!}{%
    \begin{tabular}{lll}
    \textbf{Group} & \textbf{Example Pcode Operations} & \textbf{Edge cases} \\
    \hline

    Addition/Subtraction & \pcode{IntAdd}, \pcode{IntSub} & Overflow/underflow, result negative, result zero. \\
    Multiplication/Division & \pcode{IntMult}, \pcode{IntDiv}, \pcode{IntSdiv} & Operands negative/zero. \\
    Zero/Sign Extension & \pcode{IntZext}, \pcode{IntSext} & Input all ones or all zeros, output non-negative and negative. \\
    Boolean & \pcode{BoolXor}, \pcode{BoolOr}, \ldots & Result true/false. \\
    Bitwise & \pcode{IntXor}, \pcode{IntOr}, \ldots & Result zero/non-zero. \\
    Shifts & \pcode{IntLeft}, \pcode{IntRight}, \ldots & Oversized shift operand, overflowing result, sign-bit set. \\
    Comparisons & \pcode{IntEqual}, \pcode{IntLess}, \ldots & Operands equal or $\pm1$. \\
    Overflow conditions & \pcode{IntCarry}, \pcode{IntBorrow}, \ldots &  Result true/false. \\
    Floating point & \pcode{FloatAdd}, \pcode{FloadSub}, \ldots & Operands with interesting floating point values (e.g. NaN). \\
    Branches & \pcode{Cbranch} & Condition is true and path constraints for future operations.
    \end{tabular}}
    \caption{Edge cases used for generating initial interesting input states from Pcode operations.}\label{table:emu-testing:edge-cases}
\end{table*}


\subsection{State Generation}\label{sec:emu-testing:state-generation}

The behavior of an executed instruction depends on the state of the processor. Therefore, to comprehensively test a specification, we need to explore not only changes to the instruction bytes but also to the initial input state. The state is made up of several components, including the values of ISA-level registers, memory, and processor-specific internal state (e.g., model-specific registers, caches, and micro-architectural resources). For specification testing, we assume the internal state is fixed and is reset between instruction executions. However, both ISA-level registers and memory are allowed to vary.

\noindent\scalebox{0.8}{%
\begin{minipage}{1.0\linewidth}
\begin{algorithm}[H]
  \caption{
  State Generation (from Pcode edge cases)}\label{algorithm:emu-testing:gen-input-state}
  \begin{algorithmic}[1]
    \Function{GenInputStates}{$P$} \Comment{$P$: List of Pcode operations for an instruction.}
    \State $G \gets \{\}$ \Comment{Output $G$: Set of edge case triggering input states.}
    \State $S \gets$ \textit{initial symbolic state}

    \For{$p_i$ \textbf{in} $P$}
    \State $(a, b) \gets S$.get($p_i$.inputs) \Comment{Get the current symbolic value of the inputs.}
    \State $o \gets $ \Call{Symbolise}{$a$, $p_i$.op, $b$} \Comment{Symbolically execute $p_i$.}

    \For{$e$ \textbf{in} \Call{EdgeCases}{$p_i$.op}}
    \State $s \gets $ \Call{Solve}{\Call{Assert}{$(a, b, o) \text{ triggers } e$}} \Comment{Find a concrete input $s$ triggering $e$.}
    \If{$s$ \textbf{is} \textit{SAT}}
    \State $G \gets G \cup s$
    \EndIf
    \EndFor

    \State $S$.set($p_i$.output, $o$) \Comment{Update the symbolic state with the output of $p_i$.}
    \EndFor

    \State \textbf{return} $G$
    \EndFunction
  \end{algorithmic}
\end{algorithm}
\end{minipage}
}
\vspace{2mm}

Although many specification bugs are discoverable with an arbitrary initial state, certain bugs may manifest only when an instruction is executed under extremely specific initial conditions. To allow us to test these cases, \refalgorithm{algorithm:emu-testing:gen-input-state} is used to generate interesting input states by identifying potential edge cases from the semantics section of the specification. We exploit the fact that \sle{} breaks instructions into simpler operations (Pcode) to express each edge case as an assertion in terms of the inputs and/or output of a specific operation. By systematically defining edge cases for each Pcode operation and generating initial states that trigger each edge case, it is possible to find discrepancies even when the semantics are only subtly incorrect.

\hl{\textbf{\textit{Example.}~}}Consider \hl{an \asm{ADD} instruction that also needs to update various arithmetic flags, for example on \ix{} the Pcode for \asm{ADD r0,r0,r1} is as follows}:

\begin{Verbatim}[fontsize=\small]
    OF = r0 IntCarry r1
    r2 = r0 IntAdd r1
    ZF = r2 IntEqual 0
    NF = r2 IntSignedLessThan 0
\end{Verbatim}

The \texttt{ADD} instruction is broken into four Pcode operations. Consider the second operation, \asm{r2~=~r0~IntAdd r1}, where \asm{r0} and \asm{r1} are \textit{inputs} to the instruction and are replaced with symbolic variables $a_{r0}$ and $b_{r1}$ (Line~5 of \refalgorithm{algorithm:emu-testing:gen-input-state}), the {\sc EdgeCases} method returns the following three assertions\footnote{$\sless$ represents a \emph{signed} less-than operation.}:
\begin{align*}
   a_{r0} + b_{r1} \text{ overflows}, \qquad
   a_{r0} + b_{r1} = 0, \qquad
   a_{r0} + b_{r1} \slessop 0 \qquad
\end{align*}

\noindent
Values for $a_{r0}$ and $b_{r1}$ are obtained by symbolically solving the assertions using an SMT solver (e.g., for the third assertion $a_{r0}=0, b_{r1}=-1$). After saving valid edge-case triggering inputs, the algorithm updates the symbolic state (Line~11) such that, $\texttt{r0} = a_{r0} + b_{r1}$. This means, when generating edge cases for the later Pcode operation \asm{ZF = r2 IntEqual 0}, we actually generate edge cases for $a_{r0} + b_{r1} = 0$.

All assertions are solved independently; the state generation process does not attempt to find initial input states that simultaneously solve multiple assertions. This avoids an exponential explosion of assertion combinations, many of which would be unsatisfiable. In practice, complex input states are still generated from later Pcode operations because any preceding Pcode operations that modify inputs to the operation are incorporated in symbolic expressions.
In some cases, previous operations can restrict the possible values of the operands; this makes the edge cases unsatisfiable and causes no corresponding input state to be generated.

For many ISAs, instructions may update status register bits (flags) when specific conditions occur while executing the instruction. In \sle{} specifications, these conditions are checked using individual Pcode operations, which will have edge-case-triggering inputs generated for them. This can result in the state generation process producing duplicate input states, since the conditions checked by the instruction are often the same as the edge cases defined for state generation. 

\hl{\textbf{\textit{Example.}~}}In the example code above, the same initial input state that satisfies $(r0 + r1 \slessop 0)$ (edge case for addition), also satisfies $(r2 = -1)$ (edge case for comparisons) since $r2 = r0 + r1$. However, this redundancy is useful because it ensures that missing or incorrectly implemented semantics are still tested. In cases where exact duplicates are generated, they are automatically removed before the test cases are executed.

\ifextended
    \inlinesection{Pcode Edge Cases.}
When defining edge cases for Pcode operations, there is a trade-off between defining too many, which results in excessive testing, and too few, which risks missing important behaviors. A summary of the edge cases used for state generation is provided in \reftable{table:emu-testing:edge-cases}, and we motivate them below:

\begin{itemize}[itemsep=1pt,parsep=1pt,topsep=1pt,labelindent=5pt,leftmargin=12pt]
\item \textbf{Addition and subtraction.} For \pcode{IntAdd} and \pcode{IntSub}, edge cases that test integer overflows/underflows and output values that are negative or zero are potentially interesting. Inputs corresponding to these edge cases are likely to set or clear various CPU flags, testing that the flags are properly updated by the \sle{} specification, and that the correct set of overflow checks are used.

\item \textbf{Multiplication and division.} Since there are both signed and unsigned variants of multiplication and division, edge cases are defined to identify when the incorrect variant is used. It is also useful to include edge cases for division by zero, as the behavior varies across ISAs.

\item \textbf{Zero and sign extension.} For these operations, bugs are most likely to result from using the incorrect type of extension. To identify these bugs, edge cases that ensure negative inputs of various bit-widths are used. For example, an input for a 16-bit extension would be \texttt{0xFFFF}.

\item \textbf{Boolean operations.} Given that there are only four input combinations for boolean operations, it is possible to test all outputs by constraining the inputs to each of the four combinations.

\item \textbf{Bitwise operations.} In \sle{}, bitwise operations are used for three purposes: implementing ISA-level bitwise operations, extracting and inserting bits, and as an alternative to boolean operations. Misuse of these operations can be detected by forcing the inputs to be extreme values (e.g., all bits set to 0 or 1).

\item \textbf{Shifting operations.} Overflows are also possible in shift operations, and the behavior of oversized shifts varies between ISAs. Edge cases are defined to constrain the shift operand to a value that exceeds the total number of bits in the type. Additional edge cases are defined to test whether the correct type of shift (arithmetic versus logical) is used, by constraining the inputs to be negative or positive values.

\item \textbf{Comparisons.} For comparisons, it is desirable to detect cases of off-by-one errors and where the wrong comparison operator is used. To achieve this, constraints are used to enforce that operands are equal or differ by $\pm 1$.

\item \textbf{Overflow conditions.} \sle{} explicitly calculates various overflow conditions by using specific operations for each one. Edge cases are defined corresponding to each overflow-checking operation.

\item \textbf{Floating-point operations.} For floating-point operations, the operands are constrained to extreme and special values (such as extremely small or large values, NaN, or $\pm$Inf). Floating-point operations are challenging to evaluate symbolically, which prevents us from solving subsequent edge cases that use the result of a floating-point operation. However, within a single instruction, integer and floating-point operations are typically not performed on the same inputs.

\item \textbf{Branches.} Branches are handled by forking the symbolic state at each branch operation and adding an assertion for the branch condition. For most branch operations, when the condition is true, the operation jumps to a different instruction. However, \sle{} also allows branches to Pcode operations within the same instruction (internal Pcode branches)\footnote{For example, the x86-64 specification employs internal branches to iterate over individual bits in the implementation of bit-scan instructions (\asm{BSF} and \asm{BSR}}. For the outgoing edge of external branches, edge case generation stops when the branch is taken, since instructions are individually tested. Internal loops (where the branch jumps backwards within the instruction) are handled by unrolling the loop a fixed number of times (e.g., 10) and generating edge cases for each unrolled iteration.

\item{\textbf{Memory accesses.} When a value is loaded from memory, it may be used as an input to subsequent operations. To generate edge case-triggering operands for these operations, memory is included as part of the initial input state. Whenever a load occurs, the state generator introduces a new symbolic input variable representing the value returned from the load and records the expression for the address from which the value is loaded. Since not all addresses are valid for every architecture, additional constraints are used to restrict the addresses to valid ranges and, for some architectures, to ensure that the address is properly aligned for the access size. Constraints are also added to ensure memory accesses are consistent, so that multiple loads from the same address return the same value.

For stores, the same constraints are applied to the address (to prevent invalid access exceptions from occurring), but the written value is ignored. It is technically possible for the value written by a store to be later loaded; however, there are no instructions in any of the tested ISAs where this occurs, so it can be ignored for the purpose of state generation.
}
\end{itemize}

\fi

\subsection{Differential Testing Environment}\label{sec:emu-testing:execution-environment}

\begin{table*}
\centering
\resizebox{\textwidth}{!}{%
\begin{tabular}{l|rrr|rr|c|cccc}
                                                     & \multicolumn{3}{c|}{\textbf{\textit{Generation Time (s)}}}                                                              & \multicolumn{2}{c|}{\textbf{Generated Tests}}                                   & \multicolumn{1}{r|}{{\cellcolor[rgb]{0.914,1,0.914}}}                                                                                                                                                       & \multicolumn{4}{c}{\textbf{Missing Constructors}}                 \\
\rowcolor[rgb]{0.89,0.89,0.89} \textbf{Architecture} & \multicolumn{1}{c}{\textbf{Traversal}} & \multicolumn{1}{c}{\textbf{Solving}} & \multicolumn{1}{c|}{\textbf{State Gen}} & \multicolumn{1}{c}{\textbf{Instructions}} & \multicolumn{1}{c|}{\textbf{States}} & \multicolumn{1}{r|}{\multirow{-2}{*}{{\cellcolor[rgb]{0.914,1,0.914}}\begin{tabular}[c]{@{}>{\cellcolor[rgb]{0.914,1,0.914}}r@{}}\textbf{\textit{Constructor }}\\\textbf{\textit{ Coverage}}\end{tabular}}} & \textbf{N/A} & \textbf{Redun.} & \textbf{Incom.} & \textbf{Bugs}  \\ 
\hline
\ix{}                                                  & 491.99                                 & 1019.56                              & 425.45                                  & 71,852                                    & 712,668                               & {\cellcolor[rgb]{0.914,1,0.914}}97\%                                                                                                                                                                        & 77           & 5               & 0               & 4              \\
\rowcolor[rgb]{0.89,0.89,0.89} \arch{}               & 0.24                                   & 90.09                                & 573.02                                  & 25,031                                    & 123,031                               & {\cellcolor[rgb]{0.914,1,0.914}}97\%                                                                                                                                                                        & 104          & 41              & 4               & 0              \\
\arm{}                                            & 1387.26                                & 691.99                               & 59.11                                   & 30,709                                    & 207,999                               & {\cellcolor[rgb]{0.914,1,0.914}}96\%                                                                                                                                                                        & 50           & 79              & 2               & 3              \\
\rowcolor[rgb]{0.89,0.89,0.89} \risc{}                & 0.07                                   & 39.69                                & 6.76                                    & 9,725                                     & 38,481                                & {\cellcolor[rgb]{0.914,1,0.914}}99\%                                                                                                                                                                        & 4            & 6               & 1               & 0              \\
\msp{}                                              & 0.09                                   & 46.10                                & 557.15                                  & 7,436                                    & 124,712                               & {\cellcolor[rgb]{0.914,1,0.914}}90\%                                                                                                                                                                        & 1            & 79              & 0               & 0             
\end{tabular}
}\caption{Summary of the generation process execution for each architecture. For missing constructors, \textit{N/A}: Not applicable (only used different \sle{} config), \textit{Redun.}: Redundant constructors, \textit{Incom.}: Unreferenced constructors due to incomplete specifications, Bugs: root-caused bugs in the specification.}\label{tab:testcase-generation}
\end{table*}

We use a differential testing strategy to discover bugs in \sle{} specifications using the generated test cases. The strategy is implemented in \juxtaplayer{}, which executes the test cases on two different systems: i)~an emulator derived from the \sle{} specification under test; and ii)~hardware-based reference systems.

\ifextended
    \subsubsection{System Configuration}

\inlinesection{\sle{}-based Emulator.}
\juxtaplayer{} uses \icicle{}~\cite{icicle2023}, a \sle{}-based emulator, to execute Pcode operations from the \sle{} specification. \icicle{} supports the execution of individual instructions (single-stepping) within all of the test architectures and provides APIs for directly reading and writing the system state required for differential testing. \icicle{} also supports tracking exceptions that occur during execution. 

\vspace{2mm}
\inlinesection{Hardware-Based Reference (KVM).}
Since our goal is to ensure that \sle{} specifications accurately reflect the behavior of real hardware, we use real hardware as the reference system for differential testing. Where possible, hardware-assisted virtual machines (VMs) are used to support this goal. These VMs create isolated virtual execution environments that can be used to efficiently execute instructions on the underlying hardware. By carefully configuring the VM state, it is possible to construct an environment analogous to that of the emulator.

Hardware virtualization differs between ISAs and even between different vendors of the same ISA.\footnote{Intel CPUs implement a set of virtualization extensions called VT-x, while AMD CPUs implement a different set of extensions called AMD-V (or SVM---Secure Virtual Machine).} Fortunately, the Linux kernel provides a hardware virtualization driver, KVM (Kernel-based Virtual Machine), which exposes a consistent interface for manipulating hardware VMs. KVM supports all the essential features for differential testing, allowing us to read and write register values, allocate memory, and single-step the processor.

However, some additional configuration is required to ensure that the VM environment matches the assumptions made by the \sle{} specification. This configuration is performed by \juxtaplayer{} using KVM APIs, below we describe the \ix{} specific configuration steps; the \arm{} and \arch{} configurations requires similar steps.

\ix{} processors support multiple execution modes (e.g., real mode, protected mode, and long mode), each of which can change the semantics of a particular instruction encoding. The \ix{} \sle{} specification expects to be configured to decode instructions in a specific mode. Therefore, hardware VMs must be configured to run in the correct mode (i.e., long mode for \ix{}). To do this, the relevant bits in control registers, such as \texttt{EFER} (Extended Feature Enable Register), are set to enable ISA extensions related to long mode execution, and the global descriptor table (GDT) is set up to ensure that the memory region containing the instruction bytes is interpreted as 64-bit code.

To ensure all instructions are available, \juxtaplayer{} enables additional extensions, such as SSE and AVX, by setting bits in the appropriate control registers and configuring the VM's \texttt{CPUID}. To handle memory accesses, \juxtaplayer{} also configures the page table, a hierarchical data structure used for hardware address translation, to map the instruction page and any additional memory from the test case's initial input state.

The final challenge is to support single-stepping in the VM execution environment. On both the Intel and AMD CPUs, single-stepping is implemented by setting the trap flag (TF), which causes a debug exception after each instruction returning control to the hypervisor, which in turn returns control to \juxtaplayer{}. KVM automatically handles this process, making it straightforward to single-step instructions in most cases. However, this debug exception can be suppressed by higher priority exceptions (e.g., memory violations) that occur during instruction execution. If such an exception occurs, control is not returned to the hypervisor; instead, the processor executes the corresponding exception handler. If no handler is present, a `triple fault' occurs, resetting the VM and losing the state needed for differential analysis. To prevent this, \juxtaplayer{} defines exception handlers that immediately exit back to the hypervisor. The VM is configured to use these handlers by setting the appropriate entries in the interrupt descriptor table (IDT), updating the task state segment (TSS), and mapping memory for the exception stack. When exceptions occur, we recover the exception reason using a combination of the RIP (instruction pointer) and CR3 (faulting address) registers.

\vspace{2mm}
\inlinesection{Hardware-Based Reference (Debugger).}
For \msp{} and \risc{}, we did not have access to hardware that supported hardware virtualization extensions, making it impossible to use KVM-based VMs for differential testing. In these cases, alternative approaches were required to execute instructions on real hardware while still allowing the system state to be manipulated and observed. These approaches are generally less efficient and robust than hardware virtualization, but the ISAs in question are typically simpler, making them sufficient for differential testing.

As the reference system for \msp{}, we use an MSP430FR5969 microcontroller, which contains an EEM (Energy-Efficient Microcontroller Emulator) module that can single-step instructions and read and write the system state. The microcontroller has disjoint memory regions for code, RAM, MMIO peripherals, and other functions. Gaps between these regions either trap on access or return a constant value when read.

To keep the test environment consistent, breakpoints were used to detect memory accesses outside the generic memory ranges, as no form of memory protection was available. However, only two breakpoints could be enabled simultaneously, and the breakpoint mechanism did not report information about the access that triggered it. Consequently, \texttt{NOP} instructions had to be inserted after the instruction under test for each potential trapping address, and each test case had to be executed repeatedly as part of a greater-than/less-than scan.

The lack of memory protection also meant that an instruction could write an unknown value to any MMIO register. Therefore, the microcontroller had to be fully reset between test cases to ensure a clean state.

For \risc{}, we use the BeagleV®-Ahead, which contains a quad-core Xuantie C910 processor implementing RV64GC, including the 0.7.1 draft version of the vector extension. Although the board can run Linux, it lacks support for KVM. Instead, \texttt{ptrace} was used to seize control of a newly created process and provide an isolated execution environment.

The required memory regions were mapped into the process, and the initial regions were unmapped. Because \texttt{ptrace} does not support single-stepping, test cases were first executed with memory filled with \asm{c.ebreak} instructions to detect jumps. Instructions that did not jump were then re-executed with memory populated with the values specified by the test case. Since \risc{} jump instructions do not load from memory, executions using the initial memory contents could be used as the final results for such instructions. Using \texttt{ptrace} instead of KVM limits testing to unprivileged instructions.

\subsubsection{Initial Input State}

\fi

\begin{table*}[t]
\centering
\color{black}
\resizebox{\textwidth}{!}{%
\begin{tabular}{l|l|rr|rrrrrrr}
\textbf{Architecture} & \textbf{Execs/s} & \multicolumn{1}{c}{\textbf{Raw}} & \multicolumn{1}{c|}{\textbf{Groups}} & \multicolumn{1}{c}{\textbf{Unimplemented}} & \multicolumn{1}{c}{\textbf{Alignment}} & \multicolumn{1}{c}{\textbf{Exception}} & \multicolumn{1}{c}{\textbf{Flags}} & \multicolumn{1}{c}{\textbf{Float}} &  \multicolumn{1}{c}{\textbf{PC}} & \multicolumn{1}{c}{\textbf{State}} \\ \hline
\ix{} & 176,404 & 349,897 & 18,827 & 9159 & 653 & 2467 & 3627 & 605 & 257 & 2059 \\
\rowcolor[rgb]{0.89,0.89,0.89}
\arch{} & 3,723 & 52,833 & 7,342 & 5127 & 21 & 1451 & 5 & 426 & 0 & 312 \\
\arm{} & 3,572 & 78,251 & 9,785 & 4652 & 12 & 4182 & 32 & 376 & 10 & 521 \\
\rowcolor[rgb]{0.89,0.89,0.89}
\risc{} & 1,012  & 15,682 & 937 & 322 & 0 & 373 & 8 & 222 & 1 & 11 \\
\msp{} & 0.164 & 93,043 & 2,028 & 145 & 0 & 468 & 163 & 0 & 68 & 1184
\end{tabular}%
}
\caption{
\hl{Discrepancies found by running \juxtaplayer{} on the testcases generated by \gensys{}. \textit{Raw:} total number of discrepancies (including variations to input state). \textit{Groups:} unique constructor combinations for which at least one discrepancies was found. Each of these groups was then assigned a category based on observing: the set of differences (exception state and memory/register values) and the operations in the SLEIGH semantics for the constructor combination.}}
\label{tab:discrepency-classification}
\color{black}
\end{table*}

Before an instruction is executed, the environment is configured to match the initial state defined by the test case. This includes a region of memory containing the instruction bytes with the execute permission set. Registers and other memory regions are then initialized according to the values produced by the state generation algorithm. The instruction pointer (i.e., RIP) is positioned at the instruction address, and the system is set up to execute a single step. 
The execution may succeed or raise an exception. In either event, the final system state is recorded, including modified memory values, registers, and any exception information, for later comparison. Notably, for performance, instead of mapping the entire address space,  \juxtaplayer{} lazily maps the memory whenever a memory exceptions occur and replays the instruction.

\section{Evaluation}

We focus our evaluation on five important \sle{} specifications (\ix, \arch, \arm, \risc, and \msp) to assess coverage, discrepancy patterns, and bug prevalence across diverse ISA designs. This section summarizes the results for all ISAs.

All instructions were generated using the \sle{} specification from Ghidra 10.3.\footnote{Commit: 104d9592485c2fb6c7ab5177c71d8d70140a3d03} \gensys{} is executed  on a Threadripper Pro 5995WX with 64~GB RAM\footnote{The traversal process is single threaded, while state-generation was run in parallel using 8 threads.}. Through the evaluation, we seek to answer the following questions:
\begin{itemize}[itemsep=2pt,parsep=1pt,topsep=3pt,labelindent=5pt,leftmargin=12pt]
  \item How generalizable is \name{} in traversing the ISA variants of \sle{} specifications, given the diverse specification styles used by different authors and contributors? (\refsection{test-case-generation})

  \item How effective is \name{} at finding bugs in \sle{} specifications? (\refsection{sec:emu-testing:bugs})
  
  \item What is the \hl{nature and} impact of the bugs found by \name{}? (\refsection{sec:emu-testing:bugs} and \ExtendedOrNormal{\refsection{sec:emu-testing:extra-case-studies}}{\refappendix{app:supp-materials}})

  \item \hl{How effective are the techniques employed by \name{}? (\refsection{Ablation})}
\end{itemize}

\subsection{Test Case Generation}\label{test-case-generation}

A means to objectively assess the test coverage of a given specification is necessary for assurance and evaluation of the testing method to understand its effectiveness. 
Analogous to considering code-coverage metrics as a proxy for a bug-based metric in software testing~\cite{gopinath2014,kochhar2015} or fuzzing~\cite{aflplusplus2020}, we consider measuring \textbf{constructor coverage} as a proxy for the testing coverage of a given specification. Given that the fundamental unit of syntax in \sle{} is a constructor, constructor coverage is a meaningful evaluation metric. We compute constructor coverage as:
\[
\text{Constructor Coverage} = \frac{\text{Constructors Used in Generation}}{\text{Total Constructors in a Spec.}}
\]
\noindent Further, to demonstrate the efficiency of the generation process, we also report the test \textbf{Generation Time}.

\vspace{2mm}
\noindent\textbf{Generation Results.~}\gensys{} was used to generate instructions and states for \ix{}, \arch{}, \arm{}, \risc{} and \msp{}. The results are summarized in \reftable{tab:testcase-generation}. These results represent unique constructor combinations, i.e., instructions, including variants, and input states covering the edge cases defined in \refsection{sec:emu-testing:state-generation}.

Although the instruction generation process employed by \gensys{} exhaustively explores the \sle{} specification, a total of 460 constructors remained without any generated test cases. We manually investigated each missing constructor and categorized them in \reftable{tab:testcase-generation}, in summary:

\begin{itemize}[itemsep=2pt,parsep=1pt,topsep=3pt,labelindent=5pt,leftmargin=12pt]
    \item \textbf{Not Applicable.~}Many of the missed constructors correspond to processor modes or ISA extensions that are not enabled in our configuration (e.g., \ix{} real/protected-mode semantics and \arch{} pointer-authentication support). Because these features are unavailable at runtime, the associated constructors can never be matched and can be ignored.
    \item \textbf{Redundant.~}A large proportion of the missing constructors correspond to cases where they completely overlap with a set of more specific constructors. While these redundant constructors are benign and have no impact on the specification's correctness, they can be removed to simplify the specification and improve maintainability.
    \item \textbf{Incomplete.~}Some parts of the specification are incomplete, and the missing constructors correspond to cases where a subtable constructor is fully implemented, but the parent constructor is missing.
    \item \textbf{Specification Issues (Bugs).} A small set of the missing constructors represent issues in the \sle{} specification. 
\end{itemize}

\subsection{Differential Testing Analysis}\label{sec:emu-testing:analysing-discrepancies}\label{sec:emu-testing:bugs}

Across all architectures, \juxtaplayer{} discovered 589,713 discrepancies between the output states of the \sle{}-based emulator and the hardware reference system corresponding to \totaldiscreps{} unique constructor combinations after grouping. However, many of these stem from hardware/processor-specific instructions (see \ExtendedOrNormal{\refappendix{sec:emu-testing:hardware-diff}}{\refappendix{app:supp-materials}}), or from instructions that are not fully implemented in the specification\footnote{\sle{} allows custom Pcode operations as placeholders for instructions during decompilation. Since no Pcode is associated with these operations, they cannot be emulated and analyzed using differential testing.}. Analyzing the discrepancies enabled us to identify \totalbugs{} unique bugs. In the following sections, we describe the discovered discrepancies, the analysis process and the specification bugs. We also present an analysis of root causes and sources of errors we uncovered along with bug examples. 

\subsubsection{Discrepancy Analysis and Specification Bugs}\label{sec:discrep-cats} 
In general, discrepancies correspond to either bugs or \sle{} specification limitations. In \reftable{tab:discrepency-classification}, discrepancies are first grouped by unique constructor combinations and then assigned to categories we describe here, using an ordered set of heuristics. This categorization serves two purposes. \textit{First}, it separates discrepancies that are unlikely to correspond to actionable \sle{} bugs---for example, those involving unmodeled hardware state or undefined behavior---from discrepancies that more directly indicate errors in decoding or instruction semantics. \textit{Second}, it provides a consistent link between the observed form of a discrepancy and the mechanism used to detect it, which makes the results easier to interpret.

\begin{itemize}[itemsep=1pt,parsep=1pt,topsep=1pt,labelindent=5pt,leftmargin=12pt]
    \item \textbf{Unimplemented:~}Discrepancies corresponding to either: unimplemented behavior---caused by placeholder semantic operations in the \sle{} specification; or unsupported behavior---use of architectural state not represented in our model, such as model-specific registers (MSRs) or other hardware state not defined at the ISA level. These are detected by the presence of placeholder operations (e.g., \texttt{unimpl} blocks or custom \texttt{pcodeop} definitions), or the use of unmodeled registers in the \sle{} semantics. Unimplemented behaviors often arise for recently added ISA features, or instructions with difficult semantics to directly express in Pcode. Unsupported behavior typically corresponds to hardware-specific operations that may depend on implementation-specific hardware details rather than ISA-defined semantics.

    \item \textbf{Exception:~}Discrepancies in the type or metadata (e.g., faulting address) of an exception raised by the \sle{} semantics relative to the hardware reference. It also captures cases caused by ISA feature mismatches (e.g., instructions from some ISA extensions not present on the hardware reference, but included in the \sle{} specification). Exception mismatches are also frequently related to unpredictable/undefined behavior. For example, on \arm{} some instructions can be encoded with the program counter (PC) as a  destination register, which is listed as unpredictable behavior in the architecture. The \sle{} specification treats this as ignoring the operation in many cases, while the hardware reference raises an exception. Unfortunately, the variability of exception sources makes it difficult to automatically separate out true issues in this category from unpredictable/undefined behavior.

    \item \textbf{Alignment:~}Discrepancies related to alignment issues. They are separated out as a special case of exception discrepancies since these are generally unfixable without adding additional semantics to \sle{}---see \refsection{Recommendations}.

    \item \textbf{Flags:~}Discrepancies in architectural flags, such as condition-codes or status bits. These are detected when the only observed state difference is in the flag register set. We isolate this class because flag behavior for some architectures is frequently related to undefined or unpredictable cases, and therefore benefits from a separate analysis.

    \item \textbf{Float:~}Discrepancies in registers or memory that arise along an execution path containing floating-point Pcode operations. These are detected by tracking whether the corresponding \sle{} semantics use floating-point operators. This category is separated from general state discrepancies because some differences may reflect limitations of the Pcode floating-point model rather than errors that can be fixed by modifying the \sle{} specification.

    \item \textbf{PC:~}Discrepancies in the final program counter, including incorrect branch targets or advancement to the next instruction---e.g., from decoding the wrong instruction length on \ix{} and \msp{}. These are detected when execution ends at a program counter different to the oracle. Such cases are strong indicators of decoding errors and therefore often correspond directly to \sle{} specification bugs.

    \item \textbf{State:~}Any remaining discrepancy in registers or memory not captured by the preceding categories. This is the residual class used when the final architectural state differs but no more specific heuristic applies. In practice, these cases are often the most interesting, as they may indicate incorrect semantics, missing side effects, or incorrect handling of particular operand forms.
\end{itemize}

\noindent
Notably, the categories above serve as a prioritization mechanism for analysis as much as a descriptive taxonomy since discrepancies in classes are difficult to directly associate with a specific error in the \sle{} specification. While each discrepancy corresponds to a possible bug, each bug usually results in many different discrepancies.

To analyze the discrepancies within a group, we manually aggregate discrepancies based on mnemonic, semantics, and overlapping constructors, then selected a representative example to study.
Time to triage a unique bug was approximately 4~hours on average; the time varies significantly depending on root-cause complexity. This was followed by a considerable time to craft, test and document suitable fixes for Ghidra maintainers. This process enabled us to identify \totalbugs{} unique bugs (where a bug corresponds to a  unique root-cause fixed in an ISA).

\begin{table}[t!]
\resizebox{\linewidth}{!}{
\centering
\begin{tabular}{l|cccc|c}
\textbf{Architecture} & \multicolumn{1}{l}{\textbf{Decoding}} & \multicolumn{1}{l}{\begin{tabular}[c]{@{}l@{}}\textbf{Semantics }\\\textbf{ Ordering}\end{tabular}} & \begin{tabular}[c]{@{}c@{}}\textbf{Incorrect}\\\textbf{ Semantics}\end{tabular} & \multicolumn{1}{l|}{\textbf{Aliasing}} & \multicolumn{1}{l}{{\cellcolor[rgb]{0.918,1,0.918}}\textbf{Total}}  \\ 
\hline
\rowcolor[rgb]{0.89,0.89,0.89}
\ix{}         & \hl{16} & 0     & \hl{12}    & 4     & {\cellcolor[rgb]{0.918,1,0.918}}\hl{32}\\
\arch{}       & 0       & 1     & 22    & 0     & {\cellcolor[rgb]{0.918,1,0.918}}23\\
\rowcolor[rgb]{0.89,0.89,0.89}
\arm{}        & 0     & 0     & 15    & 2     & {\cellcolor[rgb]{0.918,1,0.918}}17\\
\risc{}       & 2     & \hl{1}     & \hl{12}    & \hl{3}     & {\cellcolor[rgb]{0.918,1,0.918}}18\\
\rowcolor[rgb]{0.89,0.89,0.89}
\msp{}        & \hl{2}     & 6     & \hl{26}    & 1     & {\cellcolor[rgb]{0.918,1,0.918}}35\\ 
\hline
\textbf{Total} &  16  & 9  & 87 & 9 & {\cellcolor[rgb]{0.918,1,0.918}}\hl{125}\\                             
\end{tabular}
}
\caption{Summary of bug types by architecture.}
\label{tab:juxtaplayer-results}

\end{table}
\noindent Consequently, the bug count is not the number of buggy instructions, which would be an inflated number. Tables for the unique bugs based on our analysis of discrepancies for \ix{}, \arch{}, \arm{}, \risc{} and \msp{} ISAs are contained in \ExtendedOrNormal{\refappendix{sec:appendix}}{\refappendix{app:supp-materials}}. To characterize the types of \sle{} issues encountered across the specifications in the bugs discovered, we categorize the bugs and summarize the results in \reftable{tab:juxtaplayer-results}. We elaborate on the classes of bugs in \Circled{I-1}--\Circled{I-15} in more detail, with bug examples, and provide additional bug case studies in \ExtendedOrNormal{\refappendix{sec:emu-testing:extra-case-studies}}{\refappendix{app:supp-materials}}.

\subsubsection{Decoding Issues}\label{sec:decode-issues}
In general, decoding issues occur when incorrect constructor constraints prevent the correct instruction implementation from being selected. We detail these in the following.

\inlinesection{\Circled{I-1}~Redundant and out-of-order prefixes.~} Instruction prefixes are one of the most complex aspects of \ix{} instruction encoding and causes discrepancies between hardware vendors (see \ExtendedOrNormal{\refsection{sec:emu-testing:hardware-diff}}{\refappendix{app:supp-materials}}), security-critical CPU bugs~\cite{reptar2023}, and severe performance issues~\cite{agnermicroarch2023}. Similarly, the \ix{} \sle{} specification contains many prefix-related decoding issues. \ix{} instructions allow redundant and duplicate prefixes; however, the \sle{} implementation assumes that at most one of each prefix is present. For example, when multiple operand-size (\asm{66}) or address-size (\asm{67}) override prefixes appear, the specification computes the final address/operand size incorrectly, leading to discrepancies.

Additionally, the 64-bit extension to the ISA introduces a new class of prefixes (the REX prefix) with different decoding rules compared to legacy prefixes. According to the ISA reference, there must be at most one REX prefix present, and it must follow all legacy prefixes. However, both the Intel and AMD CPUs tested are more lenient during decoding; any REX prefixes that precede legacy prefixes are ignored, and they allow multiple REX prefixes, only using the final one. The \sle{} specification does not replicate this behavior; instead, it rejects all instructions where there are multiple or out-of-order REX prefixes.

While compilers avoid emitting unnecessary prefixes, thus reducing the impact of this issue on real-world code, such prefixes occasionally appear in handcrafted assembly for low-level performance optimizations. For example, OpenSSL uses redundant prefixes to align code to micro-architecturally friendly boundaries~\cite{openssl-blob}.

\inlinesection{\Circled{I-2}~Context handling issues.~}Historically, the \ix{} \sle{} specification used a single context field for both indicating whether long mode (64-bit) instructions were decodable and for tracking the current address size. This design assumed that long mode could be inferred from the address size, because it has a different default setting compared to other processor modes. However, including an address-size override prefix (\asm{67}) changes the current address size, causing the specification to mistakenly assume that the processor is not in long mode. Although the specification was later refined to introduce a dedicated context bit for long mode, many long mode instructions still reference the operand-size bit, resulting in discrepancies.

\inlinesection{\Circled{I-3}~Overlapping constructors.~}Some constructors are unreachable because every instruction encoding that matches them is instead matched by a different constructor (with different semantics). Further analysis revealed that many overlapping constructors are \textit{under-constrained}. This bug causes the wrong instruction semantics and disassembly to be used for the instructions associated with an unreachable constructor.

\inlinesection{\Circled{I-4}~Over constrained constructors.~}Several constructors have additional constraints that are not present in the underlying ISA. For example on \ix{}, two constructors require a specific bit in the decoder context to be matched. However, this bit is only set in 32-bit protected mode, even though the associated instructions are also valid in 64-bit long mode. This issue likely stems from a mismatch in the way context bits are used in the \sle{} specification. On real hardware, the protected mode bit is set in long mode, but the \sle{} specification only sets an equivalent bit when configured for 32-bit mode. The impact of this bug is that the corresponding instructions are incorrectly treated as invalid in long mode.

\inlinesection{\Circled{I-5}~Constructor ordering.~}For \ix{} specifically there were also several issues related to constructor ordering. Two unreachable constructors, corresponding to 64-bit variants of instructions, share the same encoding as their 32-bit counterparts, except for a mandatory \asm{REX.W} prefix. However, the 32-bit variant allows the prefix to be present and is always matched first, preventing any encoding from matching the 64-bit variant. This bug likely originates from unclear ordering rules. Since the 64-bit variant is not strictly more constrained than the 32-bit variant, the decoder defaults to the declaration order. This issue causes the wrong register sizes (32-bit instead of 64-bit) to be used for the instruction, resulting in incorrect disassembly and semantics.

\subsubsection{Semantics Ordering Issues} 

\Circled{I-6}~Semantic ordering issues arise when correct outcomes are dependent on a specific order of operations which are not followed. This was common in the \msp{} ISA which makes heavy use of addressing modes to augment its basic instructions. These modes include direct and offset indexing, and address auto incrementing among others.
All of which are available to the \asm{MOV} instruction leading to cases such as \asm{MOV.W @R7+, 0x0(R7)}.
The \msp{} was found to execute this instruction by resolving and incrementing the source register, before resolving the destination, performing the instruction's operations with the pre-incremented value, and writing back the result.
This causes loading the destination \asm{0x0(R7)} to be impacted by the increment of the source \asm{@R7+}.
This differed from the \sle{} implementation which always performed the increment as the last operation of the instruction.
This also occurred in pseudo-instructions such as \asm{POP.W SP} (\asm{MOV.W @SP+, SP}) even though they had separate Pcode semantics.

Semantic ordering issues were also found outside of addressing modes in \arch{}. The \asm{NGCS} instruction's output was written to the destination register, and then the flags were calculated off that register's value.
However when using the zero register as the destination register no write is performed, and the flags are always incorrectly calculated from the fixed value of zero.

\subsubsection{Incorrect Semantics Issues}
Incorrect or missing semantics bugs appear when an implementation uses the incorrect operations, or does not implement all behaviors of an instruction. This was the largest category of bugs, and also contained the most \textit{subtle} and unique mistakes. For example, in \arch{} the \asm{BCAX} instruction was implemented with an \textit{or} operation instead of an \textit{xor}, and \asm{STLRB} stored a word sized value instead of a byte. In addition to these subtle issues, there were several wider classes of issues.

\inlinesection{\Circled{I-7}~Missing floating point semantics.~}\risc{} floating point extensions encode smaller precision floats as NaN-boxed inside larger registers by setting all the unused upper bits to 1's.
But, the \sle{} specification does not use this encoding for single precision floats when writing to the registers, nor does it propagate un-NaN-boxed values as NaN's, causing all initial floating point test cases to fail.
The extra logic that fixing this issue added to the start and end of all floating point instructions was not able to be filtered out by \ghidra{}'s decompiler, causing the decompilation to become bloated.

\inlinesection{\Circled{I-8}~Default values and behaviors.} \risc{} defines specific result values for arithmetic exceptions. On division by zero, the division instruction should return $2^{width}-1$, and the remainder should be the dividend. There was no zero check before the division, resulting in this behavior being dependent on the execution environment default.

\inlinesection{\Circled{I-9}~Incorrect register lists.~}
To implement instructions operating on groups of registers in \sle{}, recursive subtables are used where the different constructors conditionally include the Pcode semantics for that register.
This occurs in \msp{}, \arch{}, and \arm{}, with \msp{} having an issue where it did not correctly handle register wrapping, and \arm{} having an issue where the registers were pushed in the wrong order.
However, this style of implementation inherently requires duplicated sets of constructors based on the number of registers, requiring the same fix to be applied to each constructor in the set.

\subsubsection{Aliasing Issues}\label{sec:aliasing}

\Circled{I-10}~The \sle{} specification implements instructions using multiple Pcode operations. Consequently, when source and destination operands refer to the same register or memory location (i.e., they alias), special care is required to preserve the atomicity of input values. This issue was frequently handled incorrectly in SIMD instructions (\ix{}, \arm{}), where the part of the destination register is often modified before the source register had been completely read.

For \risc{} this occurs in the Control and Status Register extension in the \asm{FSCSR}/\asm{FSFLAGS}/\asm{FSFRM} instructions. These instructions swap a value into the float CSRs by moving the source register to the CSR and the value in the CSR to the destination register. The Pcode implementation of this instruction attempted to do this without using a temporary, causing the value in the source register to be overwritten. This lack of a temporary when using aliased registers also occurs in the compressed instruction extension with the \asm{C.JALR ra} instructions. This instruction should jump to the address held in the \asm{ra} register and replace its value with the address of the following instruction, but was found to instead execute the instruction at the following address.

Both \risc{} and \arch{} exhibit bugs involving the architectural zero register, which should always return 0 and discard writes. The \risc{} \sle{} specification tries to enforce this via a subtable that exports a temporary variable instead of the zero register itself. However, several instruction definitions directly reference the zero register, circumventing this mechanism and allowing non-zero values to be written when the register is used as a destination.

Additionally for \risc{}, the destination register for some instructions is intended to be 32 bits instead of the full register width. To handle this, a constructor \asm{rdW} is used, which creates a temporary of the right size, sets its value to that of the destination register, and exports the temporary. When an instruction attempts to write to this aliased register, it would change the temporary but not the destination register. This prevented the \asm{FCVT} instructions from outputting any result.

\subsubsection{Unresolvable Issues}\label{sec:unfixed}
The problems we discussed previously can be addressed through careful editing of processor specifications. However, there are numerous issues requiring more fundamental design changes to simultaneously support accuracy of both decompilation and emulation use cases; we discuss these next.

\inlinesection{\Circled{I-11}~Unimplemented operations.~}
The SIMD and vector instructions across \ix{} and \arch{} are partially implemented in Pcode and partially with added custom Pcode ops, with the semantic section of \risc{}'s vector extension completely unimplemented.
A full Pcode implementation of these is possible, but would bloat the generated disassembly and output Pcode for static analysis tools.

\inlinesection{\Circled{I-12}~Floating point issues.~}
Floating point instructions contributed to a large number of discrepancies, as many architecture-specific aspects of the IEEE 754 standard are not exposed through the \sle{} specification layer.
\sle{} does not enforce a rounding mode and instead uses the environment's default, while modern architectures (\ix{}, \arch{}, \risc{}) provide a means to change the default rounding mode, as well as override the rounding mode for individual instructions.
Architectures with hardware support for floating point numbers generally include the common encodings of FP32 and FP64 which are supported in \sle{}. However, \ix{} and \arch{} also support the FP16 and BF16 encoding which \sle{} does not have.
\sle{} also chooses to propagate NaN payloads, while the IEEE 754 standard leaves this to the implementers' discretion. This differs from \risc{} and conditionally \arch{} which always return the canonical NaN. Properly fixing these discrepancies would require better floating point support from \sle{}. 

\inlinesection{\hl{\Circled{I-13}}~Unpredictable behavior.~}
A set of discrepancies for \ix{}, \arch{}, and \arm{} came from defined unpredictable outputs in related ISA specifications. But, these currently cannot be \textit{fixed} as there is no single correct behavior.

\inlinesection{\hl{\Circled{I-14}}~Memory alignment exceptions.~}Many SIMD and atomic operations have stricter alignment requirements than standard load/store instructions. When an operand is provided at an insufficiently aligned address, an exception should be generated. However, there is no mechanism to annotate these requirements in a \sle{} specification, meaning the emulator is unaware that an exception should occur. While extra Pcode could be used, this case is better handled by adding explicit alignment to memory accesses.

\inlinesection{\hl{\Circled{I-15}}~Missing instruction semantics.~}The x86-64 specification does not implement the auxiliary carry flag (\asm{AF}) in \textit{any} instruction, resulting in discrepancies in almost all arithmetic instructions. This is a known limitation of the specification (treated as a single bug) and likely remains unfixed because the auxiliary carry flag is rarely used in modern programs.

\rqq{Importantly, based on our in-depth analysis of the issues uncovered, we detail possible solutions in recommended modifications to the \sle{} DSL in \refsection{Recommendations}.}

\subsubsection{Discrepancy-Bug Associations}\label{sec:disc-bug}
\begin{table}[t!]
\color{black}
\centering
\resizebox{\linewidth}{!}{%
\begin{tabular}{l|c|c|c|c|c|c}
\textbf{\textit{Category}}                         & \textbf{Unimpl.} & \textbf{Exec.} & \textbf{Flags} & \textbf{Float} & \textbf{PC} & \textbf{State}  \\
\hline
\rowcolor[rgb]{0.89,0.89,0.89} Decoding            & 5                & 11             & 0              & 0              & 0           & 9               \\
Semantic Ordering                                  & 0                & 0              & 1              & 0              & 1           & 7               \\
\rowcolor[rgb]{0.89,0.89,0.89} Incorrect Semantics & 1                & 1              & 8              & 4              & 6           & 66              \\
Aliasing                                           & 0                & 0              & 0              & 0              & 0           & 10              
\end{tabular}
}
\caption{\hl{Summary of the the root-cause bugs triaged and fixed and their  association to observed discrepancies types.}}
\label{tab:disc-bug}
\end{table}
Although the relationship between discrepancies and bugs is not a simple one-to-one mapping, \reftable{tab:disc-bug} discusses how  \textit{triaged and fixed} bugs in \reftable{tab:juxtaplayer-results} relate to relevant observed discrepancies  to distinguish those that relate to the bugs. Notably, a single bug may manifest in different types of discrepancies while a single discrepancy may be the result of multiple types of bugs.
We observe up to three separate bugs triggered by a single discrepancy. Multiple discrepancies from the same bug are common, especially when incorrect semantics are encountered. Hence these numbers represent a lower bound. In addition, recall that \textit{Alignment} related discrepancies are separated since these are generally unfixable without adding additional semantics to \sle{} as we discuss in \refsection{Recommendations}. To generate results, we used the related discrepancies leading to the identification of the bug during triaging to conduct the analysis. The associations in \reftable{tab:disc-bug} demonstrate that systematic discrepancy detection serves as a strong indicator for underlying correctness problems. The distribution of bugs in \reftable{tab:disc-bug} suggests \textit{State} related issues should be prioritized in analysis.
Although  discrepancy classes are difficult to attribute directly to bugs, \textit{State} and \textit{Exception} related discrepancies are strong indicators of errors in the \sle{} specification.

\subsection{Effectiveness of Techniques}\label{Ablation}

In this section we investigate the effectiveness of the designed technique to traverse a given specification to generate encoded instructions and discover specification bugs. 

\begin{table}[b!]
\color{black}
\resizebox{\linewidth}{!}{%
\begin{tabular}{l|rc|rc|rc}
\multirow{2}{*}{\textbf{Arch.}} & \multicolumn{2}{c|}{\textbf{Simple Only}} & \multicolumn{2}{c|}{\textbf{Explicit Only}} & \multicolumn{2}{c}{\textbf{No Backtracking}} \\ \cline{2-7} 
 & \multicolumn{1}{c}{\textbf{\textit{$\Delta$~Acc}~($\downarrow$)}} & \multicolumn{1}{c|}{\textbf{\textit{Missed}}} & \multicolumn{1}{c}{\textbf{\textit{$\Delta$~Acc}~($\downarrow$)}} & \multicolumn{1}{c|}{\textbf{\textit{Missed}}} & \multicolumn{1}{c}{\textbf{\textit{$\Delta$~Acc}~($\downarrow$)}} & \multicolumn{1}{c}{\textbf{\textit{Missed}}} \\ \hline
\rowcolor[rgb]{0.89,0.89,0.89}\ix{} & 16.7\% & 995 & 12.3\% & 65 & 0\% & 0  \\
\arch{} & 13.2\% & 98 & 13.7\% & 79 & 0\% & 11 \\
\rowcolor[rgb]{0.89,0.89,0.89}\arm{} & 30.4\% & 302 & 15.9\% & 80 & 0\% & 75\\
\risc{} & 57.7\% & 38 & 54.7\% & 38 & 0\% & 0 \\
\rowcolor[rgb]{0.89,0.89,0.89}\msp{} & 69.6\% & 388 & 41.9\% & 58 & 0\% & 0 
\end{tabular}%
}
\caption{\hl{Summary of instruction generation accuracy (based on constructors used for decoding compared to generation), and missed constructors compared to GenSys, when solving techniques are removed.}}
\label{tab:generation-accuracy}
\end{table}

\inlinesection{Instruction Generation.~}To quantify the contribution of individual components in \gensys{}, we perform an ablation study over its constructor traversal and constraint-solving techniques. We focus on three key aspects to evaluate: i) \textit{Simple constraints only} (ignoring all deferred constraints); ii) \textit{Explicit constraints only} (solve deferred constraints, but excludes implied constraints from overlapping constructors); and iii) \textit{No backtracking} (solve all constraints, but assume backtracking from subtables is impossible) versus \gensys{} (solve all constraints with backtracking allowed). 

We report two metrics to understand impact: i)~$\Delta$~\textit{Accuracy}---the decrease in generation accuracy relative to \gensys{}, where accuracy is the percentage of generated instructions whose decode path matches the target constructors; and ii)~\textit{Missed}---the number of constructors for which no valid instruction encoding was found, despite a valid encoding being found by \gensys{}.

\begin{figure}[t]
    \centering
    \includegraphics[width=0.9\linewidth]{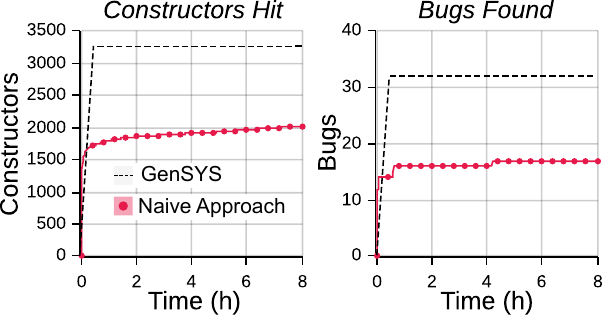}
    \caption{Constructors hit in the \ix{} \sle{} specification and bugs found with the naive approach for testing with \juxtaplayer. Notably, \gensys{} generation and the execution of the resulting test cases takes <40 minutes for \ix{}.}
    \label{fig:random-baseline}
\end{figure}

\reftable{tab:generation-accuracy} shows that simple constraint solving alone, \textit{Simple Only}, significantly reduces accuracy, particularly when applied to the \msp{} and \risc{} \sle{} specifications. Without translating simple constraints, many generated instructions match different constructors to those intended, resulting in missed constructors.
Although, \textit{Explicit constraints only}, shows a reduced accuracy drop, especially for \msp{}, accuracy drops remain above 12\% for each ISA. Then, with \textit{No Backtracking} only, all of the generated instructions' decode path matches the target constructors, as with \gensys{}, but leads to \textit{Missed} constructors. Overall, disabling the constraint translation components significantly degrades test generation effectiveness across all evaluated ISAs, causing large regions of the specification to remain uncovered.

\inlinesection{Comparison to naive approach.} To further evaluate the overall effectiveness of the instructions and states generated by \gensys{}, we can compare it to a naive approach devoid of our key techniques. Instead of modeling the \sle{} decoding process, we simply select a constructor at random, set any bits that have fixed constraints for that constructor, and fill any remaining bits with random values. For each generated instruction, we randomize the input state to derive the instruction used by \juxtaplayer{} for testing.

We compare the random approach to \gensys{} on \ix{} across two metrics: i)~\textit{Constructors Hit}---the number of constructors for which at least one generated instruction correctly decodes to that constructor; and ii)~\textit{Bugs}---the number of bugs triggered from the \xbugs{} root-cause bugs discovered for \ix{}.

Notably, fully validating bug metrics would require a significant amount of manual analysis, as indeed performed in our primary analysis with \name{}, for the randomly generated instructions. Therefore, we automatically determine whether a bug was triggered. We manually annotated each bug with the conditions required for detecting it---a \textit{canary} based on the constructors containing the bug. We provide canaries for each bug in our open-source GitHub repository.

The results of the naive approach are shown in \reffig{fig:random-baseline}. After generating and testing $1.1 \times 10^7$ testcases taking 8 hours, the naive approach failed to: i)~generate any instruction that matches 1249 constructors (38\% miss); and ii)~find 15 bugs (47\%~miss). To better illustrate the failures of the naive approach, we consider two of the missed bugs below:

\textit{\asm{CVTSS2SI/CVTSD2SI}.~}These instructions are used to convert scalar floating-point values into integers. However, the \sle{} specification contains a bug where the float-to-integer cast operation (\asm{trunc}) is missing in the constructors that handle cases where the destination operand size matches the floating-point source operand size. These instructions have a complex encoding that requires a specific prefix byte and a multi-byte opcode, consequently they were never generated by the random approach. \gensys{} generates constraints that force the appropriate prefix byte when generating this instruction and is able to find this bug.

\textit{\asm{PACKUSWB}.} This instruction converts 16-bit signed words (ranging from $-2^{15}$ to $2^{15} - 1$) in the source operand into 8-bit unsigned bytes (from $0$ to $2^8 - 1$) for the destination operand. Values outside the destination range are saturated (i.e., negative values become 0 and values greater than 255 become 255). However, the \sle{} specification contains an off-by-one error in a comparison operation, causing an input of 255 to be converted to zero. Since this bug occurs only for one specific input value, random testing is highly unlikely to expose it. In contrast, \gensys{} derives edge-case inputs from the \sle{} semantics and generates targeted states around boundary conditions, allowing it to detect the bug.

\ifextended

\subsection{Bug Case Studies}\label{sec:emu-testing:extra-case-studies}

\inlinesection{Incorrect \mspx{} Constants.} \msp{} instructions use the following word to hold 16-bit constants for indexing, absolute addressing, and immediate values. \mspx{} extends these constants to 20 bits, with the upper 4 bits sourced from the preceding extension word. 
In the \sle{} specification for \mspx{} the upper 4 bits were missing from all uses of the 20 bit constants. Preventing these instructions from representing any negative number, or positive numbers larger than $2^{16}$, including memory addresses. The effects of this can be observed in Ghidra's disassembly when moving the \asm{.data} section outside of the first 64KB of memory (\reffig{fig:msp430-bug}).

\inlinesection{\risc{} CSRRW/CSRRWI.} The \asm{CSRRW} instruction atomically swaps a value into a Control and Status Register (CSR). In the case where the zero register \asm{x0} is used as the destination register, no read should be performed, preventing any CSR read side-effects.
Currently, the \sle{} specification checks if the source register is \asm{x0} rather than the destination register. This prevents handling the case where the value of the CSR is read and then cleared, as using \asm{CSRRW rd, csr, x0} will cause the CSR to be cleared without returning any value, leading to missing data in decompilation and emulation. This appears in OpenSBI~\cite{opensbi} where the \asm{CSRRWI} instruction is used to read the \asm{MTOPEI} status register and clear it.

The CSR instructions where also the only instruction that could write non-zero values to \asm{x0}. During normal emulation and analysis this has little impact as the constant 0 is returned on \sle{} reads to \asm{x0}. However, the GDB tool queries the emulator for register values, including the now non-zero zero register. This then causes single stepping branches which compare to \asm{x0} to fail. As instead of properly single stepping, GDB places a breakpoint at the incorrectly-computed next instruction address, and continues execution. 

\inlinesection{AArch64 STRLB/STRLH.} These instructions store a byte/half-word to memory. However, the \sle{} specification did not set the size of the operation, causing the entire 32-bit register value to be written to memory. This results in bytes adjacent in memory to the target being corrupted, impacting emulation accuracy.

\inlinesection{AArch64 LD2R/LD3R/LD4R.} These instructions load multiple values from memory, and store repeating copies of each value into different SIMD registers. However, the \sle{} specification only wrote to the last register given. This results in only one value being loaded from memory and stored in a register. Disassembly or emulation based on values intended to be loaded into earlier registers would yield incorrect results.

\inlinesection{AArch64 RMIF.} This instruction sets the CPU flags based on an input value and a bit-mask. However, the \sle{} specification did not correctly filter based on the bit-mask. All masked bits would be set to the value of the NG flag. This can impact both disassembly and emulation, as many flags influencing branches could be overwritten, causing some branches to no longer be reachable.

\begin{figure}
    \centering
    \includegraphics[width=0.5\textwidth]{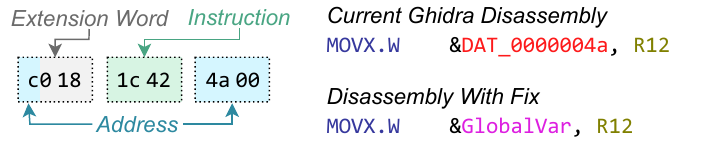}
    \caption{An example of the \msp{} bug where the upper 4 bits of 20 bit constants are ignored. This shows \ghidra{} failing to resolve the symbol as it uses the wrong address.}
    \label{fig:msp430-bug}
\end{figure}

\fi

\section{Discussion}
In addition to the limitations of this work and potential for future work, potential changes to the design of the \sle{} specification language are discussed below. These suggested changes allow for improved maintainability, versatility, fidelity, and understandability of \sle{} specifications.

\ifextended
\subsection{Recommendations for Improving \sle{}}\label{Recommendations}

The \sle{} language is demonstrably effective for specifying instruction semantics, as evidenced by the more than \totalslefams{} ISAs represented in \sle{} within \ghidra{}. However, our study reveals that \sle{} lacks features and constructs needed to express complex instruction behaviors suited to its applications in security workflows. Based on the issues uncovered through differential test-result analysis in \refsection{sec:emu-testing:bugs}, we propose extensions to the \sle{} language that could reduce bugs in specifications. We divide these extensions into two groups: i)~\textit{general recommendations}, based on the issues uncovered in \refsection{sec:decode-issues}-\ref{sec:aliasing}, that aim to better support the key use cases for \sle{} while reducing bugs during specification development and maintenance; and ii)~\textit{recommendations to address the unresolvable issues} identified in \refsection{sec:unfixed}.

\subsubsection{General Recommendations}
The modifications described here aim to improve the expressiveness and usability of \sle{}, reducing bugs during specification development and maintenance.

\vspace{2px}
\rqqtitle{rec:hints}{Add support for decompiler \textit{hints}}{Decompiler hints can enable high-fidelity \sle{} specifications to support accurate emulation while producing clear decompilation output.}

As the fidelity of \sle{} semantics increases, the readability and overall quality of the resulting decompiled code can sometimes decline. These regressions typically arise from rare edge cases that compilers guarantee cannot occur but that the decompiler cannot reliably rule out. Examples include handling offsets to 20-bit addresses on \msp{} and differences in the bit-level representation of \texttt{NaN} values. Introducing decompiler hints---such as annotations indicating that certain conditions are likely---would preserve high-fidelity Pcode for emulation while enabling the decompiler to make simplifying assumptions that improve its output. Such hints would help specifications remain both accurate for emulation and useful for decompilation.

\vspace{2px}
\rqqtitle{rec:templates}{Introduce templated constructors and arrays}{Provide higher-level language constructs that reduce code duplication and copy-and-paste errors while improving specification readability and maintainability.}\label{rec:template}

Duplicated specification logic often causes the same issue to recur across constructors. This duplication is partly necessitated by the limited expressiveness of \sle{}, particularly its inability to represent loops and array-like constructs. When an instruction requires a repeated action, such as assigning values to a list of registers (as in \reflisting{arm-reglists}), authors must write each assignment manually. This limitation not only contributes to bugs but also makes specifications harder to read and maintain. Consequently, complex instructions are sometimes left unimplemented. Extending \sle{} with arrays and templated constructors could address this problem. Possible syntax for this extension is shown in \refappendix{sec:appendix-extensions}: a template for the \texttt{reglist} constructor uses the variable \texttt{i} to index arrays containing both the constructor constraints and the target registers.

\vspace{2px}
\rqqtitle{rec:accessor}{Introduce improved accessor functions}{The current use of the \asm{export} statement to return references from subtables is error-prone. Dedicated accessor functions would provide a clearer mechanism for exporting values and managing dependencies.}
\label{rec:accessor}

To pass information between subtables, \sle{} allows a semantics section to return a reference to a value, variable, or memory location through an \asm{export} statement, which must be the final statement in the section. Subtleties in the behavior of \asm{export} make it error-prone and contribute to non-obvious aliasing problems.

A key subtlety is that an \asm{export} statement is not evaluated alongside the rest of its semantics section. Instead, it is evaluated when referenced by the semantics section of the parent constructor. Consequently, copying a register value to a temporary and exporting that temporary behaves differently from exporting the register directly. In the aliasing bugs discovered in the \ix{} specification, the main constructor modifies a register without accounting for the effect on references exported by subtables. Dedicated accessor functions could prevent such bugs by, for example, requiring the use of temporaries. Alternatively, the compiler could warn when an exported reference may overlap a register modified by the parent constructor.

\vspace{2px}
\rqqtitle{rec:helpers}{Support inline helper functions}{Inline helper functions would encapsulate complex logic for reuse and allow tools to select the desired representation and fidelity of instruction semantics.}\label{rec:helper}

\sle{} currently relies on black-box functions---e.g., \texttt{define pcodeop <name>}---for instructions that are too complex to model in Pcode or whose semantics are not fully implemented. Although these black-box functions are generally intended to be replaced eventually with Pcode semantics, the reverse sometimes occurs. For example, the Pcode semantics for the \ix{} instruction \texttt{FXSAVE} were replaced with the black-box \texttt{\_fxsave} operation, matching the compiler intrinsic used in C to generate the instruction. This representation simplifies decompilation but cannot be used for emulation.

Inline helper functions could permit dynamic selection of the desired representation and fidelity of instruction semantics. An emulator, could maximize fidelity by handling every edge case, while the decompiler could represent the same semantics concisely as a function call.

\subsubsection{Recommendations to Address Unresolvable Issues}

We identified problems, described in issues \Circled{I-11}--\Circled{I-14}, that are infeasible or impractical to resolve within the current \sle{} language. To address these limitations, we propose the following extensions to the \sle{} DSL:

\rqqtitle{rec:simd}{Support SIMD extensions \Circled{I-11}}{Expressing SIMD semantics in Pcode is cumbersome because many repeated operations are performed on different parts of the input operands. Introduce vectorized Pcode operations or structured iteration constructs to represent SIMD instruction semantics more concisely.}\label{rec:SIMD-ext}

Modern ISAs increasingly include instructions that operate on lanes of data, such as SIMD and vector instructions, yet expressing their semantics in \sle{} remains cumbersome and error-prone. Consequently, we observe poor decompilation quality and a history of bugs in the subset of these instructions that have been defined. As more programs adopt these instructions, strong support in \sle{} specifications becomes increasingly necessary. The extensive use of custom Pcode operations and the partial SIMD implementations we found suggest that the Pcode instruction set should include a new group of SIMD operations.

Inline helper functions could improve decompilation by allowing the decompiler to hide internal details. However, they may not benefit advanced decompilation techniques such as de-vectorization---i.e., translating a sequence of vector instructions back into a loop over scalar elements. New vectorized Pcode operations would make SIMD constructs easier to express, although every analysis that consumes Pcode would need to account for them. Alternatively, \sle{} could provide structured iteration constructs that are automatically unrolled for analyses requiring unstructured Pcode while remaining visible to the decompiler.

\rqqtitle{rec:floats}{More precise floating-point semantics~\Circled{I-12}}{Introduce language constructs for describing floating-point formats, operations, rounding modes, and related semantics more precisely.}

The limited set of floating-point operations in \sle{} is insufficient to capture the complete behavior of floating-point instructions in modern ISAs. For example, Pcode provides no operations for unsigned conversions between integer and floating-point values and no method for encoding different formats, such as \texttt{bfloat16} and half-precision floating point, or rounding modes.

New Pcode operations for floating-point formats, conversions, and rounding modes would enable more precise \sle{} specifications. However, every analysis that consumes Pcode would need to account for these operations, potentially requiring changes to all existing Pcode-based analyses and tools. Alternatively, some semantics could be encoded as hints, similar to existing branch hints. Tools such as decompilers could then selectively ignore the additional detail and treat an operation as a more generic case.

\rqqtitle{rec:unpredictable}{Represent unpredictable and undefined values explicitly~\Circled{I-13}}{Add mechanisms to model and isolate undefined or architecturally unpredictable ISA behavior more precisely.}

A considerable number of discrepancies across all architectures arise from outputs designated as undefined or unpredictable by the ISA reference. Such outputs can also differ across hardware implementations of the same ISA. One remedy is to add a Pcode operation that marks a register as containing an unpredictable value. This operation would allow the resulting discrepancies to be isolated more precisely, help analysis tools identify portability problems caused by compiler bugs or manually written code, and assist in analyzing malware that exploits implementation differences to identify its execution environment.

\rqqtitle{rec:alignment}{Specify memory alignment requirements~\Circled{I-14}}{Allow alignment constraints to be expressed explicitly when performing memory operations.}

Currently, \sle{} supports only a \textit{global} alignment constraint for all memory operations within an ISA. This limitation can cause discrepancies in the exception state after emulating instructions with stricter alignment requirements. Possible solutions could include: adding alignment information to the Pcode \pcode{Load}/\pcode{Store} operations or introducing naturally aligned memory operations. Operation-level alignment would improve emulation accuracy and benefit static analyses by supporting the detection of potentially unaligned accesses, improving type inference, and allowing constant-propagation algorithms to make assumptions about registers that hold aligned pointers.

\fi


\subsection{Other Processor Specifications}

\name{} is designed to address the challenges posed by the \sle{} DSL in its extensive and diverse set of processor specifications but its core ideas---symbolically traversing a specification's decoding rules, extracting edge cases and state generation---can extend to other specification languages such as Sail or VADL. At a high level, languages such as VADL~\cite{openVALDFlorian2026} and Sail~\cite{reid2016trustworthy} share structural similarities with \sle{}: they describe instruction encodings alongside executable or interpretable semantics. VADL, as implemented in OpenVADL~\cite{openVALDFlorian2026}, provides a concise and formal representation of processor architectures, with explicit modeling of instruction formats and behavior. Similarly, Sail~\cite{reid2016trustworthy} offers a rich, formally grounded language for ISA semantics, often used in verification settings for architectures such as ARM. These commonalities suggest that our approach could generalize along two main dimensions. First, specification traversal---i.e., systematically enumerating decoding paths, instruction variants, or semantic rules---can be adapted to any language that exposes structured instruction definitions. Second, symbolic solving can be applied as long as the specification can be translated into a constraint representation. In principle, both VADL and Sail provide sufficient structure and formalism to support these operations.

However, adopting our approach requires solving challenges pertinent to the specification language, and traversal in each specification language is fundamentally different because each language exposes a different normalized representation and structure. In particular, more formally grounded specification languages such as Sail support richer decoding constructors (e.g., complex control flow), which can significantly increase solver complexity or introduce path explosion during specification traversal, requiring more aggressive pruning or summarization. Then, translating semantics into efficient encodings for a symbolic solver could be non-trivial, if the abstractions do not map directly to bit-precise constraints. Hence, the primary effort lies not in redesigning the symbolic methodology itself, but in translating language-specific representations into a model suitable for constraint solving and automated exploration.

\subsection{False Positives}

In general, discrepancies correspond to either bugs or limitations in the \sle{} specification, thus defining a false positive is hard. Each discrepancy identified represents a possible bug, but cannot be easily deduplicated without manual effort, so we report discrepancies, as done in related works in CPU emulator testing~\cite{emufuzzer2009, pokeemu2012, kemufuzz2010}. Then, we discuss bugs in the context of root causes fixed, not the number of buggy instructions, a more inflated number.

The large number of discrepancies found for many of the identified bugs is a major contributor to the observed difference between the bug and discrepancy counts. In addition to discrepancies mapped to root-causes, other contributors include ISA-defined unpredictable behaviors, unimplemented semantics, or issues which cannot be fixed without extending \sle{}, discussed in \refsection{sec:unfixed}. Due to limitations of the \sle{} language, these issues do not represent an error in the specification, and consequently could be considered false positives; however, our recommendations for improving \sle{}, discussed in \refsection{Recommendations}, would allow addressing the sources of these discrepancies while helping to improve specification veracity.

\subsection{Limitations \& Future Work}
Despite our extensive efforts, our work is not without limitations. Although we sought to test the correctness of a specification, our approach will not determine the completeness of a specification, since instructions missing from a \sle{} specification are not generated, and therefore not tested. However, these instructions exhibit a well-defined failure mode: whenever an unknown instruction encoding is encountered, an invalid instruction exception will be generated in, for instance, \sle{}-based emulators. This behavior typically makes any missed instructions easy to detect in real-world targets. But, in disassemblers, missing instructions in specifications can impact control-flow recovery where invalid instructions are used as heuristics for identifying unreachable or invalid code regions. In these cases, missing instructions may cause the analysis to misattribute control-flow errors, making it harder to localize the root cause. Future work should consider not only correctness but also achieving completeness as an attribute of specification testing. 

Some ISAs include instructions that alter the execution of subsequent code. For example, to allow for a more compact instruction encoding, \arm{} includes an If-Then-Else (\asm{ITE}) instruction in \textsf{THUMB} mode that changes up to four consecutive instructions into variants that are conditionally executed based on the value of the status register. Since the approach presented here generates and tests a single instruction at a time, it does not test these cases. Therefore, future work can consider advancing the generation process to identify and generate the necessary instruction sequences to test these variants in the specification.

\section{Related Work}
We acknowledge the significant efforts devoted in past studies to test CPU emulators. EmuFuzzer~\cite{emufuzzer2009} tested \ix{} CPU emulators using a hardware reference to detect structures within the instruction by observing when mutating a particular byte causes an exception. 
KEmuFuzz~\cite{kemufuzz2010} extends EmuFuzzer by adding support for fuzzing virtual machines and hypervisors through carefully crafted templates that include snippets of both user- and system-level instructions, but retained the same instruction generation method as EmuFuzzer. The approach is suited for testing \ix{} emulators running in protected mode (32-bit instructions), requires filtering branch instructions 
due to execution environment limitations, and only identifies imprecise byte-level structure in instructions.

In contrast, PokeEMU~\cite{pokeemu2012} generates test instructions by symbolically executing the emulator's \hl{binary} instruction decoding logic. This is similar to the constructor traversal process adopted in \gensys{}, but applied to the emulator binary. 
Fast PokeEMU~\cite{fastpokeemu2018} builds on PokeEMU by reducing test case execution overhead by merging multiple instructions into a single test case. In contrast to \name{}, PokeEMU does not generate edge case-triggering input states, as the value of input operands is not used in the decoding logic. Further, \name{'s} objective to guarantee complete constructor coverage implies that the method devised can find bugs in unreachable constructors.

Other testing strategies are found in studies examining binary lifters. MeanDiff~\cite{testingir2017} compares binary lifters by translating each intermediate representation (IR) into a common IR. Once translated, an SMT solver is employed to check for symbolic equivalence. However, due to the complexity of the translation process, certain complex instructions are omitted from analysis. In contrast, \juxtaplayer{}, akin to CPU-emulator testing methods, uses a hardware oracle as a reference and determines equivalence by testing selected initial input states, avoiding the need to check the symbolic equivalence of semantically complex instructions, such as floating-point operations.
Notably, a prior study analyzed the limitation of Pcode's formal semantics~\cite{naus2022}. This led to several recommendations to improve the Pcode definitions, documentation and resolve bugs in Ghidra. This complements the recommendations resulting from our extensive work across five ISAs to improve \sle{} and the processor specifications in \refsection{Recommendations}.

Alongside these studies, we also acknowledge studies developing techniques to uncover flaws in hardware-assisted virtual machines~\cite{virtcpuvalidation2015,multinyx2018} or methods to uncover discrepancies in the textual representation of instructions generated by disassemblers~\cite{nversiondisasm2010}. Other approaches aim to detect unexpected changes in the processor state of individual hardware CPUs arising from undocumented instructions~\cite{sandsifter2017}, manufacturing defects~\cite{silifuzz2021}, or hardware implementation issues~\cite{silifuzz2021}. 

In general, similar to our study, these works across several domains also aim to identify discrepancies. Notably, discrepancies, even one known to differ between hardware implementations, can be exploited by malware for emulator or VM detection, as demonstrated in previous studies~\cite{redpills2009,cardinalpill2014,handlingantivm2017}.

\section{Conclusion}
This paper introduced a methodology for generating instructions and initial input states that fully explore all machine code sequences decodable by \sle{} processor specifications. This led to the development of \name{} with \gensys{} and \juxtaplayer{}. \gensys{} is an instruction and state generator used to produce over 1,200,000 test cases across \isacount{} different ISAs, covering all instructions and addressing modes, as well as edge cases. The test cases were used to evaluate the correctness of specifications using a differential testing strategy, realized in \juxtaplayer{}. The process revealed systematic issues in \isacount{} \sle{} specifications and highlighted areas where improvements to the \sle{} specification language could reduce future defects. Improving the veracity of \sle{} specifications enhances the reliability and confidence in downstream tools that depend on their correctness. Importantly, our testing strategy and tools  support automated correctness evaluation of \sle{} specifications into the future.

\section*{Acknowledgements}
The work was supported by the Next Generation Technology Fund (NGTF)---administered by Commonwealth Scientific and Industrial Research Organisation (CSIRO) and Defence Science and Technology Group (DSTG) in Australia, The Advanced Strategic Capabilities Accelerator (ASCA) program and DSTG, Australia. The authors would like to thank and acknowledge Om Mandavia for the preliminary investigations.

\appendix

\section{Ethical Considerations}
Our work presents a method and tool design for testing and evaluating processor specifications focusing on the open source \sle{} specifications. Our goal is to improve the truthfulness of the tools used by the security community, such as \ghidra{} and CPU emulators, that rely on the truthfulness of the \sle{} specifications, and thereby helping to improve the security of software systems and the safety and security of end-users of those systems.

Through our analysis with \name{}, we identified a substantial number of defects in existing \sle{} specifications. While documenting these issues is essential for improving the correctness and reliability of disassembly and analysis frameworks, public discussion of such defects carries some dual-use risks. In particular, an adversarial actor could potentially misuse specification inconsistencies to craft binaries that evade static or dynamic analysis, mislead emulators, or otherwise undermine defensive tooling.

To mitigate these risks, we followed established responsible-disclosure practices. All defects discovered during this work were communicated directly to the \ghidra{} project maintainers, together with the corresponding fixes that we validated internally. Our intention is to strengthen the ecosystem of analysis tools rather than expose previously unknown issues without remediation.
We emphasize that any future defects uncovered through the use of this tool should likewise be discussed and disclosed responsibly to the maintainers of the \ghidra{} project following the guidelines therein.

Despite these concerns, we believe the benefits of improving specification quality, strengthening capabilities and correctness of widely used analysis tools, employing those specifications, used by security practitioners and improving transparency outweigh the risks or potential for misuse. Ultimately, our commitment to responsible disclosure and to support the maintainers of widely used tooling guided the design of this work and will continue to guide its future development. 

\section{Open Science}
\name{} code and artifacts in the study are publicly archived at \href{https://doi.org/10.6084/m9.figshare.32507811.v1}{https://doi.org/10.6084/m9.figshare.32507811.v1}, including.
\begin{itemize} [itemsep=2pt,parsep=1pt,topsep=2pt,labelindent=5pt,leftmargin=12pt]
    \item \textbf{\gensys{}}, the main implementation of our techniques, used to traverse the \sle{} specifications and generate the test cases summarized in \reftable{tab:testcase-generation}.
    \item \textbf{\jp{}}, used for executing the test cases on both real-world hardware and in \sle{} emulators. As well as for for identifying and grouping the discrepancies for \reftable{tab:discrepency-classification}.
    \item \textbf{Documentation}, on how to use each tool in the form of multiple \asm{README} files, and on how to recreate the ablation study in \refsection{Ablation}.
    \item \textbf{Bug Fixes}, individualized in our fork of the \ghidra{} project while they undergo the review process.
\end{itemize}

We open-source the \name{} project on GitHub at \href{https://github.com/Sleigh-InSPECtor/}{https://github.com/Sleigh-InSPECtor}.

\bibliographystyle{abbrv}
\bibliography{references}

\ifextended
\section{Hardware Discrepancies}\label{sec:emu-testing:hardware-diff}

\subsection*{x86-64 Hardware Discrepancies} 

For \ix{}, we experimented with different hardware reference systems for differential testing: an \intelcpu{} and an \juxtaplayerIAcpu{} (the CPU used for the final results).
We observed many discrepancies between the two CPUs, most of which were expected. The \ix{} ISA includes instructions that expose model-specific or internal processor state, such as \asm{RDTSC}, \asm{CLZERO}, \asm{CPUID}, and \asm{RDRAND}, and therefore produce processor-dependent outputs. The processors also implement different sets of ISA extensions and vendor-specific features, in part because the AMD processor is newer than the Intel processor.

Other expected differences arise from outputs that the vendors' ISA manuals explicitly define as undefined. These cases mosly occur in arithmetic instructions for which some flags are undefined after execution, giving hardware implementations flexibility. For example, many discrepancies result from the AMD processor not setting the auxiliary carry flag (\asm{AF}), which was designed for binary-coded decimal (BCD) arithmetic and is rarely used by modern programs. We also identified four notable discrepancies that are neither attributable to ISA extensions nor explicitly documented as undefined:

\begin{itemize} [itemsep=2pt,parsep=1pt,topsep=2pt,labelindent=5pt,leftmargin=12pt]
  \item Control-flow instructions with an operand-size override prefix (\asm{66}) are interpreted differently by the two processors. The Intel CPU respects the prefix and truncates the destination address to 16 bits, while the AMD CPU ignores it and uses the full 64-bit address.

  \item Because of subtle differences in prefix decoding, the Intel CPU decodes the byte sequence \asm{[f3 47 90]} as \asm{PAUSE}, while the AMD CPU decodes it as \asm{XCHG}.

  \item The \asm{VGATHER/VPGATHER*} instruction family loads values from memory using vector index and mask operands.  Both the index and mask operands are vector registers; however, only the lower 64 bits of the mask are used. The Intel CPU fills the upper bits of the mask register with copies of the most significant bit of the original value, while the AMD CPU leaves the upper bits unchanged.

  \item \asm{RCPPS}, \asm{RSQRTPS}, and their variants compute approximate reciprocal and reciprocal square root operations, respectively. Intel and AMD CPUs use different approximation algorithms, producing different results.
\end{itemize}

The discrepancies involving nonstandard instruction encodings or unusual input values are highly unlikely to occur in real-world programs. Of the four cases, only the approximation differences are likely to affect such programs. However, since the associated instructions are explicitly documented as approximate, programmers should not rely on identical results across processors. ANonetheless, knowing these discrepancies exist is necessary to accurately assess the correctness of the \ix{} \sle{} specification. We filtered most processor-dependent discrepancies by instruction mnemonic during analysis to avoid misclassifying them as specification bugs.

\subsection*{\msp{} Hardware Discrepancies}

For some discrepancies between the hardware and the \sle{} specification, the specification was actually \textit{more} faithful to the documentation.
For example, the documentation notes that the least significant bit of the stack pointer is fixed at zero and states that the stack pointer is always even-aligned. 
This description suggests that bit 0 is hardwired to zero and that an odd value cannot be written to the register. 
However, the hardware was found to allow and to use odd stack pointer values. 
As word memory accesses are forcibly aligned to word boundaries but byte accesses are not, this behavior can cause data pushed by \asm{PUSH.B} and \asm{PUSH.W} to overlap.

The behavior of \asm{CMPA SR, SR} also differed between the hardware and documentation. 
This instruction should compare the status register with itself and set the flags based on the result, but the hardware treated the status register in the destination operand position as zero. 
The documentation states that, for non-word accesses using the register addressing mode, the status register acts as a constant-generator register. However, the observed behavior is inconsistent with the documented constant generator, which yields zero when the register appears in either operand position.

Overly permissive constraints in the \sle{} specification also allowed us to discover undocumented instructions. The MSP430X extension-word encoding occupies the range from \texttt{0x14XX} to \texttt{0x1CXX}, but the documented encoding uses only \texttt{0x14XX}, with the low byte and bit 8 ignored. The \sle{} constructor for the extension word checks only bits 11--15, ignoring bits 9--10 even though they should be zero. This allowed us to find a special case where the \asm{SXTX.W} instruction is used with the register addressing mode and bits 8--10 of the extension word set (bit 8 being the zero carry bit). Normally, \asm{SXTX.W} sign-extends a byte in the destination register to a word size value, but with this illegal encoding, it instead sign-extends a word to an address sized value. The documentation describes neither this encoding nor any instruction that sign-extends a word to address size.

\subsection*{\risc{} Hardware Discrepancies}
The \risc{} \sle{} specification includes all extensions for a given base instruction set. Because test cases are generated from this combined specification, the generator cannot distinguish extensions supported by the target hardware from those it does not support. Consequently, many tests raise illegal-instruction exceptions when the hardware attempts to execute unsupported instructions.

Another source of illegal instructions came from Control and Status Register (CSR) accesses. CSRs control processor behavior and expose status information. The \sle{} implementation defines every possible CSR address, assigning unallocated registers generic names of the form \texttt{csrXXX}. 
However, the hardware faults with an illegal instruction exception on instructions that access these CSRs as they are not implemented and have no function. The set of CSRs which can be used for testing is further reduced by the privilege level the hart is running on, as accesses to those on a higher level will also raise an illegal instruction-exception.

\onecolumn
\section{Bug Tables}\label{sec:appendix}
\begin{table*}[htp]
    \resizebox{\linewidth}{!}{%
    \begin{tabular}{llll}
    \textbf{SLEIGH Issue} & \textbf{Fix Status} & \textbf{Encoding} & \textbf{Disassembly} \\ \hline
    Memory operands decoded incorrectly with address override and REX.W prefixes. & \hl{Assigned} & {[67 43 \ldots]} & Many instructions \\
    0x82 allowed as an alternative opcode for 0x80 in long-mode (should be invalid). & \hl{\textbf{Merged} (11.2)} & {[82 f8 00]} & (invalid) \\
    64-bit SBB instruction does not sign-extend the 32-bit immediate operand. & \hl{\textbf{Merged} (11.2)} & {[4f 1d ff \ldots]} & SBB RAX,-0x1 \\
    Incorrect semantics for CMOV instructions if source $=$ destination register. & \hl{\textbf{Merged} (11.2)} & {[0f 40 00]} & CMOVO EAX,{[RAX]} \\
    Multiple SIMD instructions behave incorrectly if source $=$ destination register. & \hl{\textbf{Merged} (11.2)} & {[0f 5a c0]} & CVTPS2PD XMM0, XMM0 \\
    LZCNT instruction behaves incorrectly if source $=$ destination register. & \hl{Assigned} & {[f3 0f bd c0]} & LZCNT EAX,EAX \\
    Destination address incorrect when base address register matches source register. & \hl{Assigned} & {[0f c1 00]} & XADD {[RAX]}, EAX \\
    MOVSXD/MOVZX with 16-bit destination decoded as having a 32-bit destination. & \hl{\textbf{Merged} (11.1)} & {[66 40 63 c0]} & MOVSXD AX,AX \\
    PEXTR* instructions do not write to memory when destination is a memory operand. & \hl{\textbf{Merged} (11.2)} & {[66 40 0f \ldots]} & PEXTRB {[RDX]}, XMM0, 0x0 \\
    String instructions with a 32-bit operand size fail to zero upper bits of operands. & \hl{Assigned} & {[67 af]} & SCASD {[EDI]} \\
    Loop instructions with a 32-bit operand size fail to zero upper bits of RCX. & \hl{Assigned} & {[67 e2 0a]} & LOOP RIP+10 \\
    Instructions with repeat prefix do not set RIP correctly when RCX $=$ 1. & \hl{Pending} & {[f2 a4]} & MOVSB.REPNE RDI,RSI \\
    XCHG with 32-bit operand size fails to zero upper bits. & \hl{Assigned} & {[87 00]} & XCHG {[RAX]}, EAX \\
    CMPXCHG8B fails to zero the upper bits of 64-bit registers. & \hl{Assigned} & {[0f c7 08]} & CMPXCHG8B {[RAX]} \\
    Bits 96 to 128 not compared in CMPPS semantics. & \hl{\textbf{Merged} (11.2)} & {[0f c2 c0 00]} & CMPEQPS XMM0,XMM0 \\
    Decoding issue for MOV reg,imm8 instructions with address override and REX prefix. & \hl{Assigned} & {[67 48 b7 00]} & MOV DIL,0x0 \\
    Wrong casting operation used in CVT instructions. & \hl{\textbf{Merged} (11.2)} & {[f2 0f 2d c0]} & CVTSD2SI EAX, XMM0 \\
    REX prefix causes NOP to be decoded as a 32-bit XCHG which zeroes the upper bits. & \hl{Assigned} & {[40 90]} & NOP \\
    Destination of bit-scan instructions is modified when source operand is zero. & \hl{Pending} & {[0f bc f1]} & BSF ESI,ECX \\
    Carry flag for rotate instructions should not be set when count operand is zero. & \hl{Assigned} & {[48 d3 d0]} & RCL RAX,CL \\
    BSWAP with operand size override interpreted as 32-bit instead of 16-bit instruction. & \hl{Pending} & {[66 0f c8]} & BSWAP AX \\
    POPF with address size / operand size overrides is not decodable. & \hl{\textbf{Merged} (11.4)} & {[66 67 9d]} & POPF \\
    ModRM:reg (w) destination encoding handled incorrectly (multiple instructions). & \hl{Assigned} & {[0f c5 00 00]} & (invalid) \\
    Many instructions with a `no-prefix' constraint are decoded when prefixes is present. & \hl{Pending} & {[66 0f 37]} & GETSEC \\
    PACKUSWB is incorrect when the value to convert is exactly 0x00ff. & \hl{\textbf{Merged} (11.2)} & {[66 0f 67 c0]} & PACKUSWB XMM0, XMM0 \\
    Address size prefix not handled in bit test instructions. & \hl{Pending} & {[67 48 0f \ldots]} & BTC {[EAX]},RAX \\
    MOV instructions can reference invalid control registers. & \hl{Pending} & {[0f 20 08]} & MOV RAX, CR1 \\
    MOV to/from debug registers with REX.R prefix are not treated as invalid. & \hl{Pending} & {[44 0f 23 c0]} & (invalid) \\
    \hl{Lockable INC missing \texttt{reg\_opcode} constraint.} & \hl{Assigned} & {[fe 3f]} & (invalid) \\
    \hl{Redundant and out-of-order prefixes are handled incorrectly.} & \hl{Pending} & {[66 66 \ldots]} & Many instructions \\
    \hl{Instructions invalid in long-mode are decoded with address override prefix.} & \hl{Pending} & {[67 \ldots]} & AAA \\
    \hl{No-prefix constraint not checked in several places.} & \hl{Pending} & {[66 0f 37]} & GETSEC 
    \end{tabular}}
    \caption{Issues discovered and fixed in the \ix{} SLEIGH specification. Example instruction encodings and disassembly are provided to aid interpretation, however there are typically multiple encodings with the same issue.}\label{table:emu-testing:x86-bugs}
\end{table*}

\begin{table*}[h!]
    \resizebox{\linewidth}{!}{
    \begin{tabular}{llll}
    \textbf{SLEIGH Issue} & \textbf{Fix Status} & \textbf{Encoding} & \textbf{Disassembly} \\ \hline
Incorrect operator used in vdup. & \hl{\textbf{Merged} (12.1)} & {[10 0b e0 ee]} & vdup.8 q0,r0 \\
ldaexd behaves incorrectly when registers alias. & \hl{\textbf{Merged} (11.2)} &  {[d2 e8 f0 2b]} & ldaexd r2,r11,[r2] \\
vselgt checks the wrong condition. & \hl{\textbf{Merged} (12.1)} & {[a1 7a 7a fe]} & vselgt.f32 s15,s21,s3 \\
Incorrect constraints used in THUMB encoding of vdup. & \hl{\textbf{Merged} (12.1)} & {[a0 ee 90 0b]} & vdup.32 q8,r0 \\
VMVN operands are reversed in output. & \hl{\textbf{Merged} (12.0)} & {[e0 05 f0 f3]} & vmvn q8,q8 \\
Some instructions that modify PC do not handle the THUMB bit correctly. & \hl{Assigned} & {[c7 46]} & mov pc,r8 \\
Immediates used in thumb shift operations not decoded correctly. & \hl{Assigned} & {[5f ea 10 00]} & lsrs.w r0,r0,\#0x20 \\
ldrsh.w/ldrsb.w treat the loaded value as a pointer. & \hl{\textbf{Merged} (12.1)} & {[bf f9 00 00]} & {ldrsh.w r0,[0x10004]} \\
vst2 uses wrong registers when reglist encoding wraps. & \hl{Assigned} & {[80 e3 40 f4]} & vst2.32 \{d30,d31,d0,d1\} ... \\
Arithmetic operations with shifted immediate do not handle the carry bit correctly. & \hl{Assigned} & {[10 ea 70 00]} & ands.w r0,r0,r0, ror \#0x1 \\
PC relative encodings for FLDM*X/FSTM*X load from the wrong address. & \hl{Assigned} & {[03 fb 9f bc]} & fldmiaxlt pc, {d15} \\
PC relative encodings for STR/STRT load from the wrong address. & \hl{Assigned} & {[10 f0 8f e5]} & str pc,[0x10018] \\
Aliasing issue between index and destination register for ldrexd. & \hl{\textbf{Merged} (12.1)} & {[9f 0f b0 e1]} & ldrexd r0,r1,[r2] \\
vpush/vpop order elements in reverse. & \hl{Assigned} & {[2d ed 02 0a]} & vpush {s0,s1} \\
vmov.8/16/32 calculates register subslices incorrectly. & \hl{Assigned} & {[10 0b 40 ee]} & {vmov.8 d0[0x0],r0} \\
addrmode3 loads input as pointer instead of returning it as an address. & \hl{Assigned} & {[d8 00 cf e1]} & ldrd r0,r1,[0x10] \\
sha1su1.32/sha1su1.32 performs shift on incorrectly sized register. & \hl{\textbf{Merged} (11.2)} & {[f2 07 c7 e0]} & sha1su0.32 q8,q8,q8
\end{tabular}}
\caption{Issues discovered in the ARM32/Thumb specification. \hl{Example instruction encodings and disassembly are provided to aid interpretation, however there are typically multiple encodings with the same issue.}}\label{table:emu-testing:arm-bugs}
\end{table*}

\begin{table*}[h!]
    \resizebox{\linewidth}{!}{
    \begin{tabular}{llll}
    \textbf{SLEIGH Issue} & \textbf{Fix Status} & \textbf{Encoding} & \textbf{Disassembly} \\ \hline
    RMIF uses the wrong value for masked flags. & \hl{Assigned} & {[00 04 00 ba]} & rmif x0, \#0x0, \#0x0 \\
    SHA1H uses a shift operation instead of rotate. & \hl{\textbf{Merged} (12.1)} & {[00 08 28 5e]} & sha1h s0, s0 \\
    XAR performs an or operation instead of a xor. & \hl{Assigned} & {[1d b1 93 ce]} & xar v29.2D, v8.2D, v19.2D, \#0x2c \\
    BCAX performs an or operation instead of a xor. & \hl{Assigned} & {[d4 09 25 ce]} & bcax v20.16B, v14.16B, v5.16B, v2.16B \\
    FMLSL/FMLAL use integer instead of floating point operations. & \hl{Assigned} & {[00 40 80 0f]} & {fmlsl v0.2S, v0.2H, v0.H[0x0]} \\
    EOR3 performs an or operation instead of a xor. & \hl{Assigned} & {[00 00 00 ce]} & eor3 v0.16B, v0.16B, v0.16B, v0.16B\\
    FMLS wrong operand used in calculation. & \hl{Assigned} & {[80 51 20 5f]} & {fmls h0, h12, v0.H[0x2]} \\
    FMLAL2/FMLSL2 use the wrong register subslices & \hl{Assigned} & {[22 82 86 2f]} & {fmlal2 v2.2S, v17.2H, v6.H[0x0]} \\
    RAX1 uses a shift and or operation instead of a rotate and xor. & \hl{Assigned} & {[b4 8e 60 ce]} & rax1 v20.2D, v21.2D, v0.2D\\
    SBCS/NGCS flags incorrect when destination is XZR. & \hl{Assigned} & {[9f 03 1d 7a]} & sbcs wzr, w28, w29 \\
    LDURSB zero extends value instead of sign extending it. & \hl{Assigned} & {[e0 03 c0 38]} & {ldursb w0, [sp]} \\
    STLRB/STLRH store the wrong sized value. & \hl{Assigned} & {[e0 83 80 08]} & {strlb w0, [sp]} \\
    LDPSW loads the wrong sized value. & \hl{\textbf{Merged} (12.1)} & {[e0 77 c0 68]} & {ldpsw x0, x29, [sp], \#0x0} \\
    LDRSH/LDRSB sign-extend the result to the wrong size. & \hl{Assigned} & {[e0 03 c0]} & {ldrsh w0, [sp]} \\
    FRINT* operations perform an integer round instead of a float round. & \hl{Assigned} & {[00 40 e6 1e]} & frintn s0, s0 \\
    FMLS does not perform a vectorised operation. & \hl{Assigned} & {[80 51 20 5f]} & fmls v19.2S, v26.2S, v14.2S \\
    FMOV performs a float cast instead of a copy. & \hl{Assigned} & {[00 00 e6 1e]} & fmov w0, h0 \\
    FADDP orders input vectors incorrectly in fused vector operations. & \hl{Assigned} & {[52 15 43 2e]} & faddp v18.4H, v10.4H, v3.4H \\
    FMLA uses wrong element size for vector operations. & \hl{Assigned} & {[e7 0e 41 0e]} & fmla v7.4H, v23.4H, v1.4H \\
    LDAPR does not actually load the value from memory. & \hl{\textbf{Merged} (12.1)} & {[e0 c3 a0 f8]} & {ldapr x0, [sp]} \\
    RMIF performs a shift instead of a rotate. & \hl{Assigned} & {[cf 87 1f ba]} & rmif x30, \#0x3f, \#0xf \\
    32-bit variant of LDAR loads the wrong sized value. & \hl{\textbf{Merged} (12.1)} & {[e0 83 a0 f8]} & {ldar w0, [sp]} \\
    ld2r/ld3r/ld4r write to the wrong destination registers. & \hl{\textbf{Merged} (11.4)} & {[e0 c3 60 0d]} & {ld2r \{v0.8B, v1.8B\}, [sp]}
\end{tabular}}
\caption{Issues discovered in the AArch64 specification. \hl{Example instruction encodings and disassembly are provided to aid interpretation, however there are typically multiple encodings with the same issue.}}\label{table:emu-testing:aarch64-bugs}
\end{table*}

\begin{table*}[h!]
    \resizebox{\linewidth}{!}{
    \begin{tabular}{llll}
    \textbf{SLEIGH Issue} & \textbf{Fix Status} & \textbf{Encoding} & \textbf{Disassembly} \\ \hline
    CSR instructions can write to the zero register. & \hl{Assigned} & {[73 a0 02 34]} & CSRRS zero, mscratch, t0 \\
    FSCSR/FSRM/FSFLAGS write to the zero register. & \hl{Assigned} & {[73 90 12 00]} & FSFLAGS t0 \\
    CSRRW/CSRRWI fail to write to destination when the source is zero. & \hl{Assigned} & {[f3 12 00 34]} & CSRRW t0, mscratch, zero \\
    Incorrect value is returned on divide by zero and remainder by zero. & \hl{Assigned} & {[b3 47 00 02]} & DIV a0, zero, zero \\
    Incorrect value is returned on division overflow. & \hl{Assigned} & {[33 c5 c5 02]} & DIV a0, a1, a2 \\
    Reserved instructions in the compressed extension are not decoded correctly. & \hl{Assigned} & {[02, 40]} & (invalid) \\
    Jumps in the compressed extension fail to align PC. & \hl{Assigned} & {[82 87]} & C.JR a5 \\
    `C.JALR ra' jumps to the next instruction instead of the value in ra. & \hl{Assigned} & {[82 90]} & C.JALR ra \\
    C.ADDI variants with the zero register as the destination should decode as a NOP. & \hl{Assigned} & {[05 00]} & C.NOP \\
    FSCSR/FSRM/FSFLAGS fail to swap value when source and destination registers alias. & \hl{Assigned} & {[f3 92 12 00]} & FSFLAGS t0, t0 \\
    Floating-point instructions fail to set exceptions flags. & \hl{Assigned} & {[ .. ]} & Many instructions \\
    Floating point instructions do not support NaN boxing. & \hl{Assigned} & {[ .. ]} & Many instructions \\
    FCLASS.S/FCLASS.D are not fully implemented. & \hl{Assigned} & {[d3 12 00 e0]} & FCLASS.s t0, ft0 \\
    FCVT and FMV fails to update destination register. & \hl{Assigned} & {[53 f0 00 42]} & FCVT.d.s ft0, ft1 \\
    FMV.x.w missing sign-extension. & \hl{Assigned} & {[d3 02 00 e0]} & FMV.x.w t0, ft0 \\
    FMV.x.d writes wrong value to destination. & \hl{Assigned} & {[53 05 05 e2]} & FMV.x.d a0, fa0 \\
    FCVT.wu.s/d and FCVT.lu.s/d implements unsigned truncation incorrectly. & \hl{Assigned} & {[d3 72 10 c0]} & FCVT.wu.s t0, ft0 \\
    FCVT with integer destination fail to sign extend result. & \hl{Assigned} & {[d3 72 00 c0]} & FCVT.w.s t0, ft0
    \end{tabular}}
\caption{Issues discovered in the RISC-V specification. \hl{ Example instruction encodings and disassembly are provided to aid interpretation, however there are typically multiple encodings with the same issue.}}\label{table:emu-testing:riscv-bugs}
\end{table*}

\begin{table*}[h!]
    \resizebox{\linewidth}{!}{
    \begin{tabular}{llll}
    \textbf{SLEIGH Issue} & \textbf{Fix Status} & \textbf{Encoding} & \textbf{Disassembly} \\ \hline
    Address instructions should use special immediate instead of SR and CG. & \hl{Assigned} & {[ef 02]} & ADDA \#4, R15 \\
    SUBA carry flag is inverted. & \hl{Assigned} & {[ff 0f]} & SUBA R15, R15 \\
    Indexed addressing mode does not wrap correctly. & \hl{Assigned} & {[8f 43 10 00]} & MOV \#0, 0x10(R15) \\
    Upper bits of 20 bit registers not zeroed on smaller register writes. & \hl{Assigned} & {[0f 10]} & RRC.W R15 \\
    Upper bits of 20 bit immediates are ignored when extension word is present. & \hl{Assigned} & {[0f 18 70 12 ...]} & PUSHX.A \#0xF0000 \\
    RRUX instruction missing. & \hl{Assigned} & {[40 19 0f 10]} & RRUX.W R15 \\
    Overflow bit in RRCX not reset. & \hl{Assigned} & {[40 18 0f 10]} & RRCX.W R15 \\
    PC not word aligned after indirect branches. & \hl{Assigned} & {[c0 0f]} & BRA R15 \\
    Symbolic mode does not sign extend immediates. & \hl{Assigned} & {[10 10 fe ff]} & RRC.W SYM \\
    16 and 20 bit memory accesses are not word aligned. & \hl{Assigned} & {[2f 00 01 1c]} & MOVA \&0x1C01, R15 \\
    Sign extended immediates are not masked to 20 bit. & \hl{Assigned} & {[8f 0f ff ff]} & MOVA \#-1, R15 \\
    POPM incorrectly writes values to SP or CG. & \hl{Assigned} & {[21 17]} & POPM \#3, R3 \\
    Address sized values written to PC are restricted to lower 64KB of memory. & \hl{Assigned}  & {[00 18 70 50 ...]} & ADDX.A \#0xa, PC \\
    PC as a source in the register addressing mode has the wrong value. & \hl{Assigned} & {[cf 00]} & MOVA PC, R15 \\
    PC as a source in the index addressing mode has the wrong value. & \hl{Pending} & {[3f 00 00 00]} & MOVA 0x0(PC), R15 \\
    Signed 20 bit values are not always properly handled. & \hl{Assigned} & {[df 03]} & TSTA R15 \\
    Aliasing issue with source and destination registers in ADDA and SUBA. & \hl{Assigned} & {[ef 0f]} & ADDA R15, R15 \\
    Register indirect, and indexed addressing modes cannot access memory above 64KB. & \hl{Assigned} & {[2f 10]} & RRC.W @R15 \\
    Destination for POP.B, POP.W, and POPX instructions incorrect when SP is an input. & \hl{Assigned} & {[f1 41 00 00]} & POP.B 0x0(SP) \\
    MOV, MOVX, MOVA, and BRA @Rn+ do not correctly order operations. & \hl{Assigned} & {[3f 4f]} & MOV.W @R15+, R15 \\
    CALLA decrements SP too early. & \hl{Assigned} & {[41 13]} & CALLA SP \\
    POPX does not jump when destination operand is PC. & \hl{Assigned} & {[00 18 70 41]} & POPX.A PC \\
    RRAM.W and RLAM.W compute negative flag incorrectly. & \hl{Assigned} & {[5f 02]} & RLAM.W \#1, R15 \\
    Extension word instructions that write to the CG register should be invalid. & \hl{Assigned} & {[83 00 0a 00]} & (invalid) \\
    PUSHM pushes wrong value for PC and SP. & \hl{Assigned} & {[11 15]} & PUSHM \#2, SP \\
    PUSHM does not correctly handle wrapping. & \hl{Pending} & {[f5 15]} & PUSHM.W \#16,R5 \\
    SXT does not sign extend up to 20 bits. & \hl{Assigned} & {[8f 11]} & SXT R15 \\
    The 20 bit indirect auto increment addressing mode increments by the wrong value. & \hl{Assigned} & {[00 18 7f 10]} & RRCX.A @R15+ \\
    PUSHX incorrectly orders SP register decrement. & \hl{Assigned} & {[00 18 41 12]} & PUSHX.A SP \\
    RRCX.A uses the wrong mask value. & \hl{Assigned} & {[00 18 4f 10]} & RRCX.A R15 \\
    RRAX and SXTX, and RRCX and SWPBX instruction encodings overlap. & \hl{Assigned} & {[40 18 cf 11]} & (invalid) \\
    RLAX with rpt prefix does not handle PC as the destination register correctly. & \hl{Assigned} & {[00 18 40 50]} & RLAX.A PC \\
    rpt instructions are invalid in non-register addressing modes. & \hl{Assigned} & {[40 18 13 12]} & PUSHX \#1 \\
    Variations of the address instructions do not jump when the destination is PC. & \hl{Assigned} & {[e0 03]} & INCDA PC \\
    Writes to the status register and operation flags are ordered incorrectly. & \hl{Pending} & {[02 10]} & RRC.W SR \\
    \hline
       \end{tabular}}
    \caption{Issues discovered in the MSP430X specification. \hl{Example instruction encodings and disassembly are provided to aid interpretation, however there are typically multiple encodings with the same issue.}}\label{table:emu-testing:msp430-bugs}
\end{table*}

\onecolumn
\section{Bug Tables}\label{sec:appendix}

\clearpage
\section{\sle{} Extensions}\label{sec:appendix-extensions}
\begin{listing*}[h]
  \includegraphics[width=1.0\textwidth]{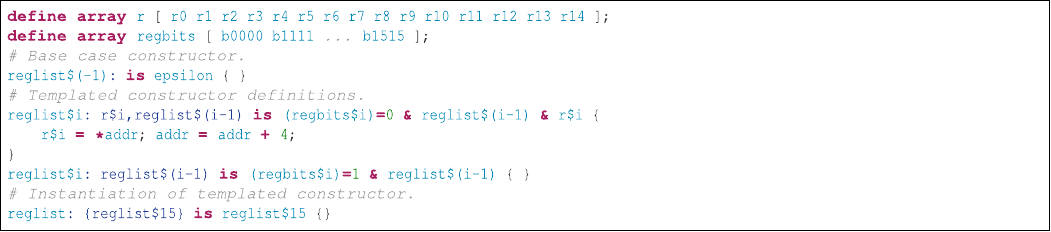}
  \caption{Example syntax for an extension to the \sle{} language that allows arrays and templated constructors to be defined.}\label{listing:emu-testing:sleigh-arrays}
\end{listing*}

\fi

\end{document}